\documentclass[aps,preprint]{revtex4}
\usepackage{amsfonts}
\usepackage{amsmath}
\usepackage{amssymb}
\usepackage{graphicx}

\input{tcilatex}
\begin{document}

\title{\textbf{Autler-Townes Multiplet Spectroscopy}}
\author{Fazal Ghafoor}
\affiliation{Department of Physics, COMSATS Institute of Information Technology,
Islamabad, Pakistan}

\begin{abstract}
We extend the concepts of the Autler-Townes doublet and triplet
spectroscopy to quartuplet, quintuplet and suggest linkages in
sodium atom in which to display these spectra. We explore the
involved fundamental processes of quantum interference of the
corresponding spectroscopy by examining the Laplace transform of
the corresponding state-vector subjected to steady coherent
illumination in the rotating wave approximation and
Weisskopf-Wigner treatment of spontaneous emission as a simplest
probability loss. In the quartuplet, four fields interact
appropriately and resonantly with the five-level atom. The
spectral profile of the single decaying level, upon interaction
with three other levels, splits into four destructively
interfering dressed states generating three dark lines in the
spectrum. These dark lines divide the spectrum into four spectral
components (bright lines) whose widths are effectively controlled
by the relative strength of the laser fields and the relative
width of the single decaying level. We also extend the idea to the
higher-ordered multiplet spectroscopy by increasing the number of
energy levels of the atomic system, the number of laser fields to
couple with the required states. The apparent disadvantage of
these schemes is the successive increase in the number of laser
fields required for the strongly interactive atomic states in the
complex atomic systems. However, these complexities are naturally
inherited and are the beauties of these atomic systems. They
provide the foundations for the basic mechanisms of the quantum
interference involved in the higher-ordered multiplet
spectroscopy.
\end{abstract}

\date{Today}
\maketitle

\section{Introduction}

The spontaneous emission spectrum of a two-level atom in the absence of any
driving field exhibits a Lorentzian line shape whose width according to the
Weisskopf-Wigner theory is proportional to Einstein's decay rate, $\Gamma $
\cite{Book1}. The spectrum modifies in the presence of a driving field. A
two-level atom leads to a three-peak resonance fluorescence spectrum when it
decays from an upper level to a lower one \cite{Book1}. However, a
three-level atom, when the driving field couples the upper two levels and
decay takes place from an intermediate level to the ground, exhibits an
Autler-Townes doublet in the spectrum \cite{Autler}. These modifications in
the spectrum arise due to Stark splitting of the driven atomic energy levels
where the decaying atom from the two dressed states interfere destructively
to create a Fano type dark line in the single Lorentzian peak \cite{Fano}. A
large number of studies exist to understand different aspects of spontaneous
emission of this type of atom-field interaction. For example in Ref. \cite%
{Agarwal} if the upper two levels are degenerate then the spectrum
can be modified because of the mechanisms of population trapping
and the quantum interference while in Ref. \cite{Agassi} if the
upper two levels are closely placed comparable to $\Gamma $ then
the single dark line always exists even in the absence of the
driving field. However this condition was avoided by Zhu
\textit{et al.} \cite{Zhu-Narducci} as the interference mechanism
is always there even in the presence of the field at any strength.
The Autler-Townes doublet and its spontaneous emission version are
based on the simple probability loss. However, the spectroscopy
both in absorption and emission context is studied extensively in
the context of naturally existing complications. Furthermore,
Paspalakis \textit{et al.} \cite{Knight01} utilized this scheme
for spontaneous emission from a coherently prepared and
microwave-driven doublet of potentially closely excited states to
a common ground. They showed that the effect of the relative phase
between the pump and the coupling field allows us to control
efficiently the spontaneous emission spectrum and the
population dynamics. Following the work of Toschek and coworkers \cite%
{Physik,Keil,Keil00} there have been a number of studies of Autler-Townes
splitting in probe absorption, when a strong laser pump field drives a
coupled transition in a vapor cell or atomic beam \cite%
{Pic,Ph,Bjor,Gray,Delsart,Fisk}. The Autler-Townes doublet scheme has been
utilized to study other areas of quantum optics. These include, for example,
the direct measurement of the state of electromagnetic fields \cite%
{Autler-Zubairy}, reconstruction of the Wigner function of non-classical
fields such as a Sch{\small \"{o}}dinger-cat state \cite{Wigner-Zubairy},
the measurement of the Wigner function for a generalized entangled state
\cite{Zubairy-Manzoor} and the measurements of atomic position in a standing
wave field \cite{Herkommer}.

The phenomenon of the Autler-Townes doublet can further be generalized to a
triplet when a four-level atom interacts appropriately with three coherent
fields \cite{Fazal}. The interaction splits the upper excited bare energy
state into a triplet. The atom now finds itself in three decaying dressed
states resulting in the quantum interference effect. Two dark lines appear
which split the single Lorentzian peak into the three components (bright
lines). These results achieved were analytical for a simplified version of
the atom-field system \cite{Fazal00} and were numerical for an experimental
feasible scheme \cite{Fazal}. Further, this scheme has been utilized for
precision measurement of atomic position in the standing wave-field from the
emission \cite{Ghafoor,Ghafoor00,F-Ghafoor} and absorption spectrum \cite%
{Sehrai}. Furthermore, the efficient control of the widths of the three
spectral components can be used for the measurement of the state of the
radiation field under relaxed experimental condition as compared to the one
reported in Ref. \cite{Autler-Zubairy}. This type of atom-field interaction
also leads to subluminal and superluminal behavior of the probe absorbed
field under the condition of EIT \cite{Sehrai1}.

Owing to the potential use of the Autler-Townes doublet and triplet
spectroscopy in the spontaneous emission study and to their applications in
other areas of quantum optics, the question arises, is it an end to the
Autler-Townes spectroscopy regarding its multiplicity? We answer this
question. We propose schemes for its generalization from the doublet and the
triplet to quartuplet, quintuplet, and consequently to the higher-ordered
multiplet spectroscopy. For the conceptual development of the mechanisms
involved in these multiplet Autler-Townes spectroscopies we need to extend
the concept to the schemes of five-level, six-level, seven-level and so on
to multi-level atomic systems interacting appropriately with the required
fields. To understand the fundamentals of these processes ideally we
examined the Laplace transform for the state-vector of these multi-state
systems subjected to steady coherent illumination in the rotating wave
approximation and Weisskopf-Wigner treatment of spontaneous emission as a
simple probability loss. The results obtained under this way are very
simple, readable and analytically interpretable. The Lorentzian line-shaped
spectrum then splits into five, six and larger numbered spectral components
associated with five, six and large-numbered decaying dressed states of the
bare energy state created by the atom-field interactions, respectively.
Initially we propose atomic schemes avoiding parity violation which can be
found in the hyperfine structured sodium atom \cite{Mirza}. However, for
higher-ordered multiplet spectroscopy, the system becomes complicated and
some of the decay processes may not be avoided. These may alter the ideal
behavior of the system but the underlying physics we expect to be
unaffected. For example, in Autler-Townes doublet and triplet schemes the
complete dark lines feature gradually becomes insignificant with the
increase in the other decay rates of the system \cite{Fazal,Zhu-Narducci}.
However, the general behavior of the system agrees with its ideal behavior.
Further, we also discuss various aspects of experimental feasibility of our
models based on the fine structure sodium D1 line adjustable with a current
experiment \cite{Xia,Zanch}. Here the only disadvantage is the complex
atomic systems. However, these complexities are naturally inherited and the
thing is to explore them for understanding the fundamental of the quantum
interference processes responsible for these multiplicity in the spectra.

The paper is structured as follows. In section II we discuss sodium atom as
a suitable experimental candidate for the realization of the different
results of Autler-Townes multiplet spectroscopy. In section III, we extend
the idea of the Autler-Townes doublet and triplet spectrum to the
Autler-Townes quartuplet spectrum and calculate the analytical expression
for it to discuss the basic mechanism involved in the interference processes
in the atomic system. Its experimental feasibility and also its
adjustability with the current technology are discussed in this section. In
the next section IV we propose another scheme to go a step beyond to
demonstrate the Autler-Townes quintuplet spectrum. Its experimental
feasibility is also discussed in this section. In section V, we present the
ways how to generalized the spectroscopy to the sextuplet and consequently
to the higher-ordered ones. Following the generalization we underline some
expected complications. We also provide the details of how to handle these
higher-ordered multiplet spectra mathematically and numerically. The section
VI is devoted to the detail discussion of our main results. Finally, the
conclusion is presented at the end of this paper.

\section{The Sodium Atom and Our Proposals}

The atomic sodium Na$^{\text{23}}$ is a alkali metal and is hydrogenoid
having no complications like the atoms of multiple valence electrons. Due to
its versatility properties, the sodium atom is famous in laboratories.
Sodium atom is very sensitive to strong fields and is useful in nonlinear
studied. It is therefore considered a well suited candidate for use in
research laboratories, particularly, in spectroscopic studies. Further, the
spectroscopic studies can be carried in the homogenous-broadened limit with
an atomic beam or a magneto-optic trap (MOT), and the Doppler-broadened
features can be observed in a vapor cell or a heatpipe.

The energy levels structure of the sodium atom in its hyperfine energy
levels in the weak magnetic field is very much compatible with our various
proposals for Autler-Townes multiplet atomic schemes. Therefore we based the
experimental proposals for quartuplet, quintuplet and consequently
higher-ordered multiplet atomic schemes on the naturally available various
excited states of the sodium atom. Furthermore, we also take the support of
a very important experiment conducted recently for multi-wave mixing by Zuo
\textit{et al.} \cite{Zanch} based on a sodium atom. In the experiment, they
achieved eight-wave mixing in sodium atom having a folded five-level
configuration in its fine structure. They suggested that their experiment is
important for the future work and might be helpful for the coherent
transient spectroscopy, Autler-Townes spectroscopy and EIT. Their method may
also particularly provide new insight into the nature of highly excited
states. Following the facilities of multiple fields coupling with the sodium
atom we underline the ways that how their experimental set-up is feasible
and adjustable in the context of our proposed schemes if based on hyperfine
energy levels of Sodium D1 line for recording the spontaneous emission
spectra. These feasibilities and adjustability are discussed in each section
for each atomic scheme under the investigation. Although we are interested
in the experiment on the wave mixing in the context of spontaneous emission
study, there are many other recently reported works \cite%
{Yanp,Zhiq,Yig,Chang,Yanp01,Chang01} related to this subject may be
interesting for future studies of Autler-Townes multiplet spectroscopy.

\begin{figure}[t]
\centering
\includegraphics[width=3.5in]{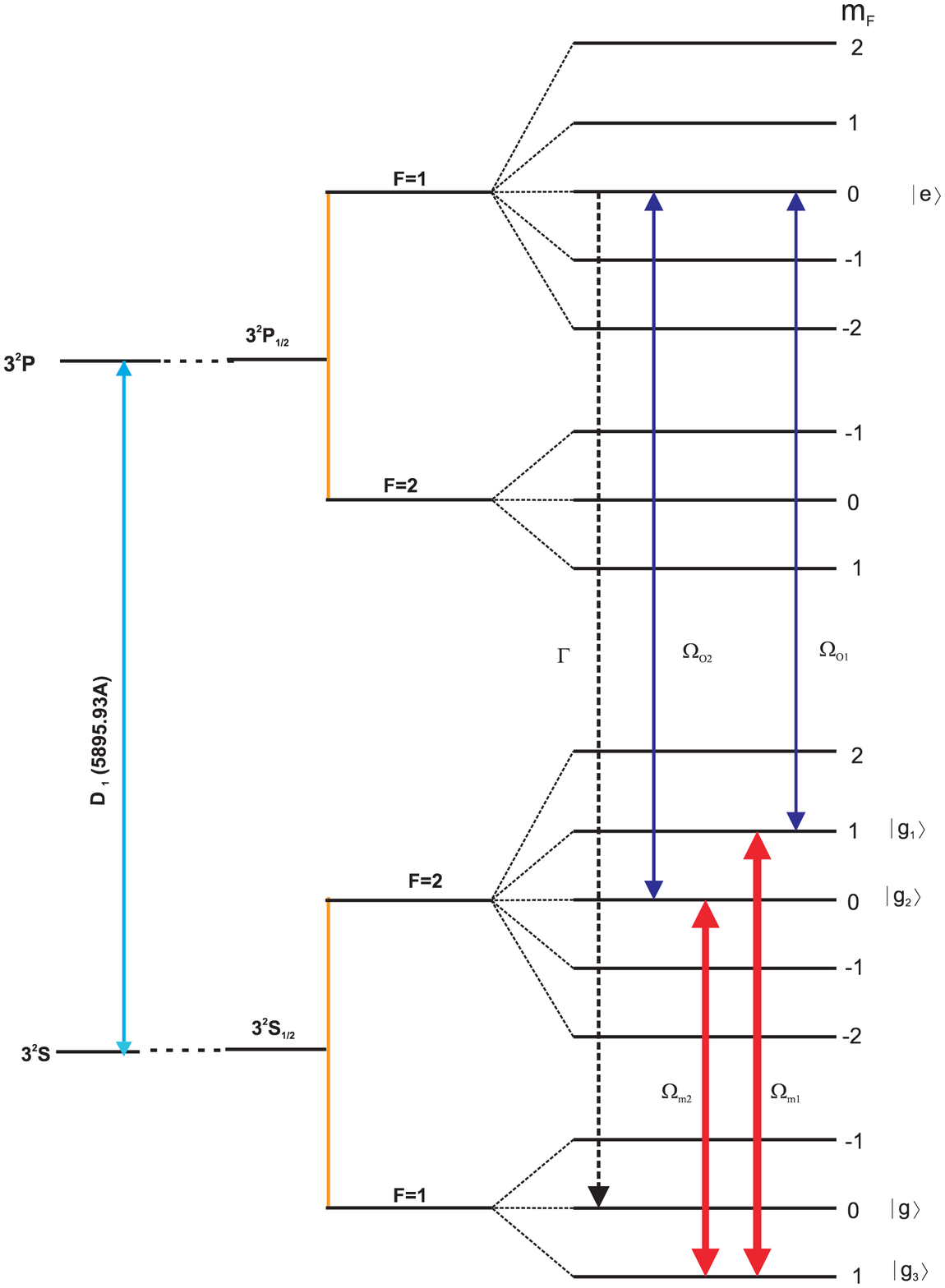}
\caption { Linkages for quartuplet spectrum of four fields and
four strongly-coupled levels of the Sodium atom. Spontaneous
emission from state $\left\vert e\right\rangle,$ shown as a dashed
linkage, is responsible for the spectrum \label{figure1}}
\end{figure}

\section{The Autler-Townes quartuplet Spectroscopy}

We propose a five-level atom with four as ground states hyperfine quadruplet
and one excited state driven resonantly by two coherent and two microwave
fields appropriately (see Fig. 1). Two pairs i.e., $\left\vert
g_{1}\right\rangle $---$\left\vert g_{3}\right\rangle $ and $\left\vert
g_{3}\right\rangle $---$\left\vert g_{2}\right\rangle $ of the four ground
hyperfine energy states $\left\vert g_{1}\right\rangle $, $\left\vert
g_{2}\right\rangle $, $\left\vert g\right\rangle $, $\left\vert
g_{3}\right\rangle $ are driven by two microwave fields having Rabi
frequencies $\Omega _{m_{1}}$ and $\Omega _{m_{2}}$ respectively. Similarly
the excited state $\left\vert e\right\rangle $ is coupled with the ground
states $\left\vert g_{1}\right\rangle $ and $\left\vert g_{2}\right\rangle $
by two coherent driving fields having Rabi frequencies $\Omega _{o_{1}}$ and
$\Omega _{o_{2}}$. The system forms the shape of loop structured
configuration under these steps. Furthermore the level $\left\vert
e\right\rangle $ is also coupled with the level $\left\vert g\right\rangle $
via vacuum field modes. The coupling of the vacuum field modes only to this
particular transition is reasonable, as in proposed configuration all
transitions from the other states to $\left\vert g\right\rangle $ are dipole
forbidden. The naturally existing one decay processes makes our system the
simplest ones due to the simply probability loss. This will leads to some
ideal behavior regarding the quartuplet spectroscopy which is necessary for
defining the fundamental of this system. More later, at the end of this
section we match this scheme with current experiments that may capable us to
realize the results of this spectroscopy in a laboratory. We also consider
the resonant atom-field interaction therefore, the interaction picture and
the rotating wave picture coincide for the coherent couplings. The
Hamiltonian under the interaction picture for the rotating wave and dipole
approximation is now given by

\begin{equation}
\mathcal{H}(t)=\mathcal{H}_{A}+\mathcal{H}_{B}\left( t\right) ,  \label{1}
\end{equation}%
where%
\begin{align}
\mathcal{H}_{A}& =\hbar (\Omega _{o_{1}}\left\vert e\right\rangle
\left\langle g_{1}\right\vert +\Omega _{o_{2}}\left\vert e\right\rangle
\left\langle g_{2}\right\vert +\Omega _{m_{1}}\left\vert g_{1}\right\rangle
\left\langle g_{3}\right\vert +\Omega _{m_{2}}\left\vert g_{2}\right\rangle
\left\langle g_{3}\right\vert )+H.C.,  \label{2} \\
\mathcal{H}_{B}\left( t\right) & =\hbar \sum\limits_{k}g_{k}e^{i\Delta
_{k}t}\left\vert e\right\rangle \left\langle g\right\vert a_{k}+H.C.
\label{3}
\end{align}%
In Eq. (\ref{3}), $g_{k}$ is the coupling constant between the $k$th mode
and the atomic dipole moment between level $\left\vert e\right\rangle $ and
level $\left\vert g\right\rangle $ while $\Delta _{k}=\omega _{eg}-\nu _{k}$
is the detuning of the $k$th mode with respect to the central frequency.
Moreover, $a_{k}$ and $a_{k}^{\dagger }$ are the annihilation and creation
operators, respectively. The statevector of the system at any time $t$ is
given by%
\begin{equation}
\left\vert \Psi (t)\right\rangle =[E(t)\left\vert e\right\rangle
+G_{1}(t)\left\vert g_{1}\right\rangle +G_{2}(t)\left\vert
g_{2}\right\rangle +G_{3}(t)\left\vert g_{3}\right\rangle ]\left\vert
\mathbf{0}\right\rangle +\sum\limits_{k}G_{k}(t)\left\vert g\right\rangle
\left\vert 1_{k}\right\rangle ,  \label{4}
\end{equation}%
where $\left\vert \mathbf{0}\right\rangle $ denotes the absence of photons
in all vacuum modes and $\left\vert 1_{k}\right\rangle $ represents a single
photon in the vacuum modes. In the infinite volume limits and apart from the
proportionality factor the spontaneous emission spectrum can be obtained as
(see Appendix-A),
\begin{equation}
S(\Delta )=\frac{\left\vert g\right\vert ^{2}\Delta ^{2}(\Delta
^{2}-\left\vert \Omega _{m_{1}}\right\vert ^{2}-\left\vert \Omega
_{m_{2}}\right\vert ^{2})^{2}}{(\Delta ^{4}-\Delta ^{2}\left\vert \Omega
\right\vert ^{2}+\left\vert \Omega _{o_{2}}\right\vert ^{2}\left\vert \Omega
_{m_{2}}\right\vert ^{2}+\left\vert \Omega _{o_{1}}\right\vert \left\vert
\Omega _{o_{2}}\right\vert \left\vert \Omega _{m_{1}}\right\vert \left\vert
\Omega _{m_{2}}\right\vert \sin \varphi )^{2}+\frac{\Gamma ^{2}}{4}(\Delta
^{2}-\left\vert \Omega _{m_{1}}\right\vert ^{2}-\left\vert \Omega
_{m_{2}}\right\vert ^{2})^{2}}.  \label{5}
\end{equation}%
Here we have replaced the set of discrete mode frequency $\nu _{k}$ by a
continuum-field frequency $\omega ,$ with that we have replaced the discrete
set of detnings $\Delta _{k}$ by continuum variable $\Delta =\omega _{eg}-$ $%
\omega .$ We have also taken the continuum to have a uniform coupling to
state $\left\vert d\right\rangle $ and have therefore replaced $g_{k}$ by a
mode-dependent $g$ (not similar with the state $\left\vert g\right\rangle $)$%
.$ Also, in the Eq. (\ref{5}), $\left\vert \Omega \right\vert =$ $\sqrt{%
\left\vert \Omega _{o_{2}}\right\vert ^{2}+\left\vert \Omega
_{m_{1}}\right\vert ^{2}+\left\vert \Omega _{m_{2}}\right\vert ^{2}}$
appears for the effective Rabi-frequency of the system. The spectrum
vanishes when $\Delta =0,$ $\pm \sqrt{\left\vert \Omega _{m_{1}}\right\vert
^{2}+\left\vert \Omega _{m_{2}}\right\vert ^{2}}$, yielding three dark lines
in the spectrum dividing it into four spectral components. The height of all
the spectral components are the same and can be calculated from the
expression $\frac{4\left\vert g_{k}\right\vert ^{2}}{\Gamma ^{2}}$. During
the derivation of the spectrum relation we assumed that the atom is
initially prepared in the excited state $\left\vert e\right\rangle $. This
way of handling the problem is mathematically correct for the ideal
behavior. However, physically it might be difficult to place an atom in its
excited state as soon as the atom enters the cavity to interact with other
driving fields. Thus from experimental point of view we ought to add the
techniques or the mechanism of the process to prepare an atom in its excited
state. This can be done very easily by coupling another field with the
ground energy level and the level where the atom is required to prepare.
This act will then bring the system into the context of resonance
fluorescence where the results obtained under strong field limit will not
match with the expected results of our system. However, the results obtained
will match under weak field limit, in which the added field strength must be
comparable with the decay rates, a well known phenomenon of resonance
fluorescence in quantum optics. Hence the assumption of initial state
preparation is reasonable and experimentally incorporable. In our system we
selected a particular nature of the interaction of the driving fields with
the atom to form a loop preserving the phase effect in the system. Each
single phase, when treated separately, has a similar effect. However, if two
or more of the phases are varied relatively, the result obtained may not
match with the ones in which the phases vary separately. In our analysis we
deal only with one phase for its effect on the system.

To support the physical aspect regarding the dressing of bare energy state
into multiple components we define the state function of the system in terms
of dressed state basis having eigen states $\left\vert \alpha \right\rangle ,
$ $\left\vert \beta \right\rangle ,$ $\left\vert \gamma \right\rangle $ and $%
\left\vert \delta \right\rangle $ with their corresponding eigenvalues $\xi
_{1}$, $\xi _{2}$, $\xi _{3}$ and $\xi _{4}$, as%
\begin{eqnarray}
\left\vert \Psi (t)\right\rangle  &=&[\alpha _{k}\left( t\right) \left\vert
\alpha \right\rangle +\beta _{k}\left( t\right) \left\vert \beta
\right\rangle +\gamma _{k}\left( t\right) \left\vert \gamma \right\rangle
+\delta _{k}\left( t\right) \left\vert \delta \right\rangle ]\left\vert
0\right\rangle   \notag \\
&&+\sum_{k}G_{k}(t)\left\vert g\right\rangle \left\vert 1_{k}\right\rangle ,
\label{6}
\end{eqnarray}%
where%
\begin{align}
\mathcal{H}_{A}\left\vert \alpha \right\rangle & =\hbar \xi _{1}\left\vert
\alpha \right\rangle ,  \label{7} \\
\mathcal{H}_{A}\left\vert \beta \right\rangle & =\hbar \xi _{2}\left\vert
\beta \right\rangle ,  \label{8} \\
\mathcal{H}_{A}\left\vert \gamma \right\rangle & =\hbar \xi _{3}\left\vert
\gamma \right\rangle ,  \label{9} \\
\mathcal{H}_{A}\left\vert \delta \right\rangle & =\hbar \xi _{4}\left\vert
\delta \right\rangle .  \label{10}
\end{align}%
The eigenvalues appearing in the Eqs. (\ref{7}-\ref{10}) are the roots of
characteristic equation of this system while the dressed states can be found
by the usual way. These eigenvalues associated with the dressed states are
also the roots of $D\left( \Delta _{k}\right) $ of the denominator of Eq.
(A6) mentioned in the Appendix-A, using the bare energy state-vector
approach. Thus the dressed statevector approach coincides with the bare
statevector at least for the association of the number of bright spectral
lines in the emission spectrum with the dressed states defined for this
system (see Fig. 2). Satisfactorily, we will use the terminology of dressed
states frequently for the explanation of our main results obtained by the
bare statevector approach. The detailed analysis for the physical processes
of the system in terms of dressed states basis is beyond the aim of this
paper and will be discussed elsewhere. However, the analysis of our main
results using bare statevector explains almost all the underlying physics of
our system.

Next we continue with our bare statevector approach. To explore many facts
about the mechanism involved in the phenomenon of quantum interference among
the four decaying dressed states, we rewrite the above Eq. (5) in the
following form%
\begin{equation}
S(\Delta )\varpropto \sum_{i=1}^{4}\left\vert \frac{g^{\ast }}{\digamma }%
\frac{\xi _{i}(\xi _{i}^{2}-\left\vert \Omega _{m_{1}}\right\vert
^{2}-\left\vert \Omega _{m_{2}}\right\vert ^{2})\zeta _{i}}{\Delta -\xi _{i}}%
\right\vert ^{2},  \label{11}
\end{equation}%
where the factor $\digamma $ and the factors $\zeta _{i}$ $\left(
i=1-4\right) $ are given in the Appendix-A. $\xi _{i}$ $(i=1-4)$ as
discussed earlier, are the roots of the following equation%
\begin{align}
& \Delta ^{4}-\Delta ^{2}\left\vert \Omega \right\vert ^{2}+\left\vert
\Omega _{o_{2}}\right\vert ^{2}\left\vert \Omega _{m_{2}}\right\vert
^{2}+\left\vert \Omega _{o_{1}1}\right\vert \left\vert \Omega
_{o_{2}}\right\vert \left\vert \Omega _{m_{1}}\right\vert \left\vert \Omega
_{m_{2}}\right\vert \sin \varphi  \notag \\
& +i\frac{\Gamma }{2}\Delta (\left\vert \Omega _{m_{1}}\right\vert
^{2}+\left\vert \Omega _{m_{2}}\right\vert ^{2}-\Delta ^{2})=0.  \label{12}
\end{align}%
The analysis of the Eqs. (\ref{11},\ref{12}) leads us to interpret the
phenomenon of quantum interference among the decaying dressed states in a
more clear and transparent way. Since, the spontaneous emission intensity is
proportional to the steady state of the absolute square of the probability
amplitude $G_{k}(t\rightarrow \infty )$ (see Appendix A) in continuum limit.
The continuum limit of the spectrum at detuning $\Delta ,$ which is now
proportional to $\left\vert G(\infty )\right\vert ^{2}$ contains four
absolute squared terms and twelve interference terms. The four absolute
squared terms can be associated with four emission probabilities from the
four dressed states of the bare energy state created by the atom-field
interaction. The interference terms are contributed by the pathways among
the upper excited four decaying dressed energy states. We note that the
interference terms compensate the contribution from the absolute squared
terms at the location $\Delta $\ = $0,$ $\pm \sqrt{\left\vert \Omega
_{m_{1}}\right\vert ^{2}+\left\vert \Omega _{m_{2}}\right\vert ^{2}}$. The
spectrum becomes zero at these values leading to three dark lines in the
total spectrum.

For appropriate choices of different spectroscopic parameters, the Eqs.
(11)-(12) get the following shape%
\begin{equation}
S(\Delta )=\Gamma \sum_{i=1}^{4}\left\vert \frac{\alpha _{i}+i\beta _{i}}{%
\Delta -(\alpha _{i}^{\prime }+i\beta _{i}^{\prime })}\right\vert ^{2},
\label{13}
\end{equation}%
where in the Eq. (\ref{13}), $\alpha _{i}$, $\beta _{i}$, $\alpha ^{\prime
}{}_{i}$, and $\beta ^{\prime }{}_{i}$ $(i=1-4)$ are the appropriate integer
values obtained due to specific choices of spectroscopic variables i.e., all
the Rabi-frequencies and the decay rate in Eqs. (\ref{11},\ref{12}). This
equation is totally the numerical version of Eq. (10). Eq. (11) consists of
four terms. Each term further consists of a numerator and a denominator.
Each numerator and denominator if subjected to the selected variables of the
system for the analysis of some behavior of the system then these terms
appear as a complex numbers with the real and imaginary parts as real
values. Thus as whole the spectrum appears as shown by Eq. (13). Here for
each set of the system variables we estimates the FWHM of each spectral
components. These estimation for various data are displayed in the form of
numerical values in the discussion parts showing the validity of the
Weisskopf-Wigner theory. These integer values appearing in the Eq. (13) are
different for the different sets of these spectroscopic variables chosen for
exploration of some behavior in the system. Now, we can readily estimate the
peaks height, their locations and widths from the absolute squared terms.
The spectrum therefore consists, in general, of four peaks locating at $%
\delta =\alpha _{i}^{\prime }$ with peak heights $(\alpha _{i}^{2}+\beta
_{i}{}^{2})/\beta _{i}^{\prime 2}$ for $i=1-4$, respectively.

\begin{figure}[t]
\centering
\includegraphics[width=3.5in]{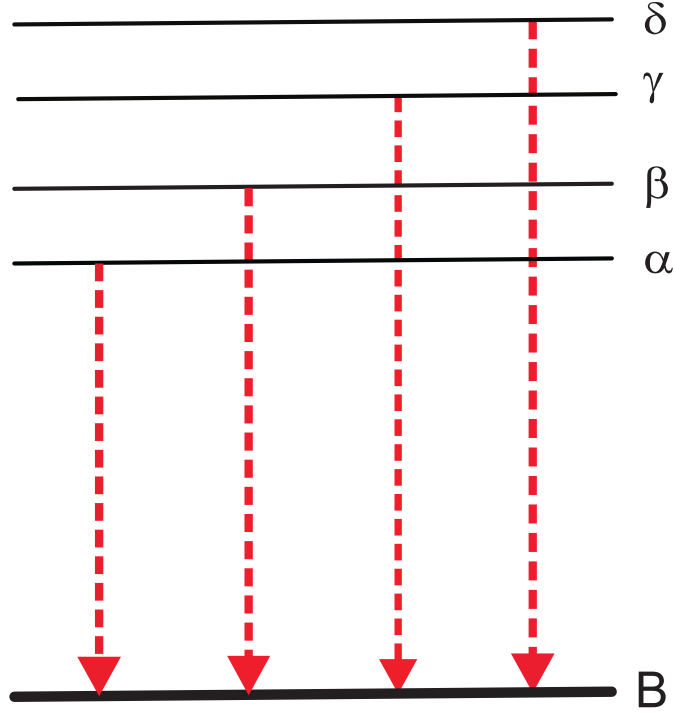}
\caption { Schematic diagram of dressed-state energies and
spontaneous emission transitions (dashed arrows) from these to the
ground state. The dressed states are evenly spaced only when all
the Rabi frequencies are equal. The dressed states of this figure
are those associated with Fig. 3 frame (a). The corresponding
dressed state diagram for the quintuplet spectrum Fig. 5 frames
\textbf{(a)} and \textbf{(b)} has five equidistant dressed states
and five spectral lines \label{figure1}}
\end{figure}

Next we discuss the experimental feasibility of our loop structured
five-level atomic scheme. We select the Sodium D1 line having two fine
structured energy levels. Each state further splits into two hyperfine
energy levels according to the usual coupling of the nuclear magnetic moment
associated with the Sodium nuclear spin and the internal atomic magnetic
field by the total electronics angular momentum of the orbital spinning
electron. For each hyperfine level, the F quantum number further splits into
$m_{F}=(2F+1)$ hyperfine sub-energy-level due to the Earth magnetic field of
strength 1 Gauss. Therefore, the Sodium $D1$ line has a total of 16
sub-energy level lines with 12 allowed transitions due selection and
polarization rules of Zeeman lines of a dipole transition. The hyperfine
Zeeman splitting within each hyperfine state of the $D1$ line all lies in
lower-frequency range of Radio frequency (RF) range i.e., $0.5$ $\sim $ $2.8$
MHz. However the typical value of Rabi frequency of a pulsed laser range
from microwave ($\mathbf{\sim }$10 MHz) to Infrared range (1 THz). The
values of the splitting of the hyperfine in the excited state is $189$ MHz
while for the ground state it is of order of $1772$ MHz, a significantly
large splitting. Both are in the range of lower frequency side of the
microwave. Therefore, it is reasonable to couple the ground hyperfine energy
levels with microwave fields. The excited allowed Zeeman energy levels of
the $3P_{1/2}$ state can then be coupled with the grounded hyperfine Zeeman
energy levels by optical frequency ranged laser fields. The simplification
of the sixteen lines of the Sodium $D1$ into four hyperfine states have also
been proposed in literature \cite{Fry,Fry01}. The arguments employed are
depended on linear polarization rules and the selection rules of the
magnetic dipole allowed transitions. Extending to five-level atomic we
select the four ground states of D1 lines i.e., $\left\vert
3S_{1/2},F_{2}=2,m_{F}=1\right\rangle \Leftrightarrow \left\vert
g_{1}\right\rangle ,$ $\left\vert 3S_{1/2},F_{2}=2,m_{F}=0\right\rangle
\Leftrightarrow \left\vert g_{2}\right\rangle ,$ $\left\vert
3S_{1/2},F_{1}=1,m_{F}=1\right\rangle \Leftrightarrow \left\vert
g_{3}\right\rangle $ and one excited state $\left\vert
3P_{1/2},F_{1}=1,m_{F}=0\right\rangle \Leftrightarrow \left\vert
e\right\rangle .$ The states $\left\vert
3S_{1/2},F_{2}=2,m_{F}=1\right\rangle \Leftrightarrow \left\vert
g_{1}\right\rangle $ and $\left\vert 3S_{1/2},F_{2}=2,m_{F}=0\right\rangle
\Leftrightarrow \left\vert g_{2}\right\rangle $ are coupled with the state $%
\left\vert 3S_{1/2},F_{1}=1,m_{F}=1\right\rangle \Leftrightarrow \left\vert
g_{3}\right\rangle $ by two microwave fields while they are coupled with the
excited decaying state $\left\vert 3P_{1/2},F_{1}=1,m_{F}=0\right\rangle
\Leftrightarrow \left\vert e\right\rangle $ by two optical fields. The
linkage of the excited state $\left\vert
3P_{1/2},F_{1}=1,m_{F}=0\right\rangle \Leftrightarrow \left\vert
e\right\rangle $ is considered with the ground state, $\left\vert
3S_{1/2},F_{1}=1,m_{F}=0\right\rangle \Leftrightarrow \left\vert
g\right\rangle $ via vacuum field modes. It is noteworthy to mention that in
our scheme the order of energy levels does not matter as we considering
resonant atom fields interaction. The linkages of the closely spaced
hyperfine states by microwave fields are usual in literature. For example,
the upper two closely spaced decaying levels of the model scheme of quantum
beat laser are prepared in coherent superposition by a microwave field. In
our system the one linkage of the optical linkages is dipole forbidden
following the selection rules of the Zeeman splitting. However, there are
experimental and theoretical methods where the linkages between the dipole
forbidden transition may be made allowed by applying small statics magnetic
of strength up to $1.2$ $mT$. For examples, the dipole forbidden transition
in the Alkaline earth elements can be allowed by the linkage of this low
statics electric field with appropriate levels, an experimental fact \cite%
{Barber}. Moreover, in the model scheme of Correlated Spontaneous Emission
(CEL) Laser the dipole forbidden states are coupled with a strong coherent
field to prepare the system in coherent superposition of the upper and the
ground states of the cascaded three-level atomic system. However, the
involvement of one decay process in this quartuplet system makes it an ideal
ones to have ideal behavior for defining the fundamentals of even this
complicated spectroscopy. This simply loss system is very much important as
there is no simple mechanism in literature (to the best of my knowledge) to
control the spontaneous decay in a system. Moreover, the multiple ground
states implementation for an atomic system are considered both theoretically
\cite{copy} and experimentally \cite{Xia}. Next we discuss the experimental
adjustability of our scheme with a very important experiment conducted
recently for multi-wave mixing reported in \cite{Zanch}. In the experiment
they achieved eight-wave mixing in sodium atom having a folded five-level in
the fine structure sodium atom. They suggested that their experiment is
important for the future work and might be helpful for the coherent
transient spectroscopy, Autler-Townes spectroscopy and EIT. Having the
facilities of the linkages of multiple fields with the sodium atom it will
be easy to go from fine to hyperfine Zeeman splitting in the Sodium D1 line
due the technological facilities of the multiple-field couplings. We are
required to link two microwave fields and two optical fields with the
selected energy levels of the sodium D1 line for our system. The initially
prepared excited state is then allowed to decay steadily to record the
spectrum.

Having quantum interference as a basic processes in the Autler-Townes
multiplet spectroscopy we make a distinction with the coherent population
trapping (CPT), a phenomenon also arising of the same effect in the
spontaneous emission studies \cite{Zhu-Scully,Paspalakis,Xia,Mart}. The area
under the curve of spontaneous emission spectrum in the case of CPT is
always less than the case when there is no quantum interference in the
system. However, here in the Autler-Townes spectroscopy, there are dark
lines in the spectrum arising because of quantum interferences among the
decaying dressed states. The area under the curve in this case is almost the
same as the one obtained from the Lorentzian line shaped spectrum associated
with a decaying bare energy state without its dressing. This fact is also
evident from our multiplet schemes where the sum of the full width at half
maximum (FWHM) of all the spectral components are always almost equal to
Einstein's decay rate, $\Gamma $, a decay rate that is normally associated
with a decaying bare energy state without its dressing. The area under the
curves of the spontaneous emission spectrum which representing the energy
given out by the atom to environment are same in the case of Autler-Townes
spectrum but not in the CPT case. This means some population is always
trapped in an upper excited dressed energy state in CPT while almost no
population trapping is there in Autler-Townes multiplet spectroscopy. In
both the CPT and Autler-Townes multiplet spectra there is quantum
interference but their approaches are different regarding its effect. In the
later case there are dark lines in the spectrum of Lorentzian line-shape
while in the former there is complete elimination (and not control) of a
spectral component.

\section{The Autler-Townes Quintuplet Spectroscopy}

Next, we modify our previous scheme for Autler-Townes quintuplet
spectroscopy. These modifications can be carried in the previous
system with a view to have a loop structured atom fields
interaction. However, the needed single looped atom fields
interaction is not unique and the problem can also be dealt for at
least one loop in order to preserve the phase effect in the
system. Therefore the higher odd-ordered multiplet schemes can be
handled through branches and loops if and only if there is no
parity violation. In the present scheme, we select one branch and
one loop to preserve the phase effect. Therefore, we consider a
six-level atom having five hyperfine ground energy states. The
three pairs $\left\vert g_{1}\right\rangle $---$\left\vert
g_{3}\right\rangle $, $\left\vert g_{3}\right\rangle
$---$\left\vert g_{2}\right\rangle $ and $\left\vert
g_{2}\right\rangle $---$\left\vert g_{4}\right\rangle $ of the
four ground states are coupled resonantly with three microwave
fields having the Rabi frequencies $\Omega _{m_{1}}$, $\Omega
_{m_{2}}$ and $\Omega _{m_{3}}$ respectively. Similarly the ground
states $\left\vert g_{1}\right\rangle $ and $\left\vert
g_{2}\right\rangle $ are coupled resonantly with the excited state
$\left\vert e\right\rangle $ by two optical driving fields having
the Rabi frequencies$\ \Omega _{o_{1}}$and $\Omega _{o_{2}}$
respectively. The structure of this atom field interaction form a
branch and a loop configurations (see Fig. 3). We expect same
results if the structure of the atom-field interaction is kept in
one loop shape. However, we are following the branch and the loop
structure to have the different possibility of getting almost the
same type results. Moreover, to include the unique allowed decay
process in the system the level $\left\vert e\right\rangle $ is
coupled with the level $\left\vert g\right\rangle $ via vacuum
field modes. Later, at the end of this section we discuss the
possibility of utilizing the experiment of Ref. \cite{Zanch} for
this quintuplet scheme and give a proposal of its adjustability
for the realization using hyperfine D1 line of Sodium. The
interaction picture Hamiltonian in the dipole and rotating wave
approximation is then given by

\begin{equation}
\mathcal{H}(t)=\mathcal{H}_{A}+\mathcal{H}_{B}\left( t\right) ,
\end{equation}%
where%
\begin{align}
\mathcal{H}_{A}& =\hbar (\Omega _{o_{1}}\left\vert e\right\rangle
\left\langle g_{1}\right\vert +\Omega _{o_{2}}\left\vert e\right\rangle
\left\langle g_{2}\right\vert +\Omega _{m_{1}}\left\vert g_{1}\right\rangle
\left\langle g_{3}\right\vert +\Omega _{m_{2}}\left\vert g_{3}\right\rangle
\left\langle g_{2}\right\vert  \notag \\
& +\Omega _{m_{3}}\left\vert g_{2}\right\rangle \left\langle
g_{4}\right\vert ))+H.C., \\
\mathcal{H}_{B}\left( t\right) & =\hbar \sum\limits_{k}g_{k}e^{i\Delta
_{k}t}\left\vert e\right\rangle \left\langle g\right\vert a_{k}+H.C.
\end{align}%
The state-vector of the system at any time t is given by%
\begin{align}
\left\vert \Psi (t)\right\rangle & =[E(t)\left\vert a\right\rangle
+G_{1}(t)\left\vert g_{1}\right\rangle +G_{2}(t)\left\vert
g_{2}\right\rangle +G_{3}(t)\left\vert g_{3}\right\rangle +  \notag \\
& +G_{4}(t)\left\vert g_{4}\right\rangle ]\left\vert \mathbf{0}\right\rangle
+\sum\limits_{k}G_{k}(t)\left\vert g\right\rangle \left\vert
1_{k}\right\rangle ,  \label{15}
\end{align}%
where $\left\vert \mathbf{0}\right\rangle $ denotes the absence of
photons in all vacuum modes and $\left\vert 1_{k}\right\rangle $
represents a single photon in the vacuum modes. In the infinite
volume limit and apart from the proportionality factor the
spontaneous emission intensity for this quintuplet atomic system
is proportional to the steady state of the absolute square of the
probability amplitude $G_{k}(t\rightarrow \infty )$ [see Eq. (B9)
of the Appendix A] in continuum limit. The continuum limit of the
spectrum $S(\Delta )$ at detuning $\Delta ,$ which is now proportional to $%
\left\vert G(\infty )\right\vert ^{2}$ is obtained as,%
\begin{equation}
S(\Delta )=\frac{\left\vert g\right\vert ^{2}\Delta ^{4}\left[ \Delta
^{2}-\left\vert \Omega _{A}\right\vert ^{2}+\left\vert \Omega
_{B}\right\vert ^{4}\right] ^{2}}{(\Delta ^{5}-\Delta ^{3}\left\vert \Omega
\right\vert ^{2}+\Delta \left\vert \Omega _{C}\right\vert ^{4}+2\left\vert
\Omega _{D}\right\vert ^{5}\cos \varphi )^{2}+\frac{\Gamma ^{2}}{4}\Delta
^{4}\left[ \Delta ^{2}-\left\vert \Omega _{A}\right\vert ^{2}+\left\vert
\Omega _{B}\right\vert ^{4}\right] ^{2}},  \label{16}
\end{equation}%
where%
\begin{align}
\left\vert \Omega _{A}\right\vert ^{2}& =\left\vert \Omega
_{m_{1}}\right\vert ^{2}+\left\vert \Omega _{m_{2}}\right\vert
^{2}+\left\vert \Omega _{m_{3}}\right\vert ^{2},  \label{17} \\
\left\vert \Omega _{B}\right\vert ^{4}& =\left\vert \Omega
_{m_{2}}\right\vert ^{2}\left\vert \Omega _{m_{3}}\right\vert ^{2},
\label{18} \\
\left\vert \Omega _{C}\right\vert ^{4}& =\left\vert \Omega
_{o_{1}}\right\vert ^{2}\left\vert \Omega _{m_{2}}\right\vert
^{2}+\left\vert \Omega _{m_{1}}\right\vert ^{2}\left\vert \Omega
_{o_{2}}\right\vert ^{2}+\left\vert \Omega _{m_{2}}\right\vert
^{2}\left\vert \Omega _{m_{3}}\right\vert ^{2}+\left\vert \Omega
_{m_{3}}\right\vert ^{2}\left\vert \Omega _{o_{2}}\right\vert ^{2},
\label{19} \\
\left\vert \Omega _{D}\right\vert ^{5}& =\left\vert \Omega
_{o_{1}1}\right\vert \left\vert \Omega _{m_{1}}\right\vert \left\vert \Omega
_{m_{2}}\right\vert \left\vert \Omega _{m_{3}}\right\vert \left\vert \Omega
_{o_{2}}\right\vert ,  \label{20} \\
\left\vert \Omega \right\vert & =\sqrt{\left\vert \Omega _{o_{1}}\right\vert
^{2}+\left\vert \Omega _{m_{1}}\right\vert ^{2}+\left\vert \Omega
_{m_{2}}\right\vert ^{2}+\left\vert \Omega _{m_{3}}\right\vert
^{2}+\left\vert \Omega _{o_{2}}\right\vert ^{2}},  \label{21}
\end{align}%
Again, we have replaced the set of discrete mode frequency $\nu _{k}$ by a
continuum-field frequency $\omega ,$ for this system. Consequently we have
replaced the discret set of detnings $\Delta _{k}$ by continuum variable $%
\Delta =\omega _{eg}-$ $\omega .$ We have also taken the continuum to have a
uniform coupling to state $\left\vert g\right\rangle $ and have therefore
replaced $g_{k}$ by a mode-dependent $g.$ Further, by inspection we note
that the spectrum is vanished at the locations:
\begin{equation}
\Delta =\pm \frac{1}{\sqrt{2}}\sqrt{\left\vert \Omega _{A}\right\vert ^{2}-%
\sqrt{\left\vert \Omega _{A}\right\vert ^{4}-4\left\vert \Omega
_{B}\right\vert ^{4}},}  \label{22}
\end{equation}%
and%
\begin{equation}
\Delta =\pm \frac{1}{\sqrt{2}}\sqrt{\left\vert \Omega _{A}\right\vert ^{2}+%
\sqrt{\left\vert \Omega _{A}\right\vert ^{4}-4\left\vert \Omega
_{B}\right\vert ^{4}}}.  \label{23}
\end{equation}%
Thus, yielding four dark lines in the spectrum dividing it into five
spectral components. The height of all the spectral components are same and
can be calculated from $\frac{4\left\vert g_{k}\right\vert ^{2}}{\Gamma ^{2}}%
.$

Again to explain the physical aspects regarding the dressing of bare energy
state into multiple components we define the state function of the system in
terms of five dressed-states basis having eigenstates $\left\vert \alpha
\right\rangle $, $\left\vert \beta \right\rangle $, $\left\vert \gamma
\right\rangle $, $\left\vert \delta \right\rangle $ and $\left\vert \epsilon
\right\rangle $ with their corresponding eigenvalues $\xi _{1}$, $\xi _{2}$,
$\xi _{3}$, $\xi _{4}$ and $\xi _{5}$, respectively%
\begin{align}
\left\vert \Psi (t)\right\rangle & =[\alpha _{k}\left( t\right) \left\vert
\alpha \right\rangle +\beta _{k}\left( t\right) \left\vert \beta
\right\rangle +\gamma _{k}\left( t\right) \left\vert \gamma \right\rangle
\notag \\
& +\delta _{k}\left( t\right) \left\vert \delta \right\rangle
+\epsilon _{k}\left( t\right) \left\vert \epsilon \right\rangle
]\left\vert 0\right\rangle +\sum_{k}G_{k}(t)\left\vert
g\right\rangle \left\vert 1_{k}\right\rangle ,  \label{24}
\end{align}%
where%
\begin{align}
\mathcal{H}_{A}\left\vert \alpha \right\rangle & =\hbar \xi _{1}\left\vert
\alpha \right\rangle ,  \label{25} \\
\mathcal{H}_{A}\left\vert \beta \right\rangle & =\hbar \xi _{2}\left\vert
\beta \right\rangle ,  \label{26} \\
\mathcal{H}_{A}\left\vert \gamma \right\rangle & =\hbar \xi _{3}\left\vert
\gamma \right\rangle ,  \label{27} \\
\mathcal{H}_{A}\left\vert \delta \right\rangle & =\hbar \xi _{4}\left\vert
\delta \right\rangle ,  \label{28} \\
\mathcal{H}_{A}\left\vert \epsilon \right\rangle & =\hbar \xi _{5}\left\vert
\epsilon \right\rangle .  \label{29}
\end{align}%
The eigenvalues appearing in the Eqs. (\ref{25}-\ref{29}) are the roots of
the characteristics equation of this system and are also the roots of $%
D\left( \Delta _{k}\right) $ appeared in the Eq. (B9) of the Appendix-B
while these dressed states can be calculated by the usual way. For this
system the dressed statevector approach also coincides with the bare
statevector at least for the association of the five bright spectral lines
in the emission spectrum with the five dressed states defined for this
system. Again, we are safe to use the terminology of the dressed states for
the explanation of the results obtained under the bare statevector approach.
\begin{figure}[t]
\centering
\includegraphics[width=3.5in]{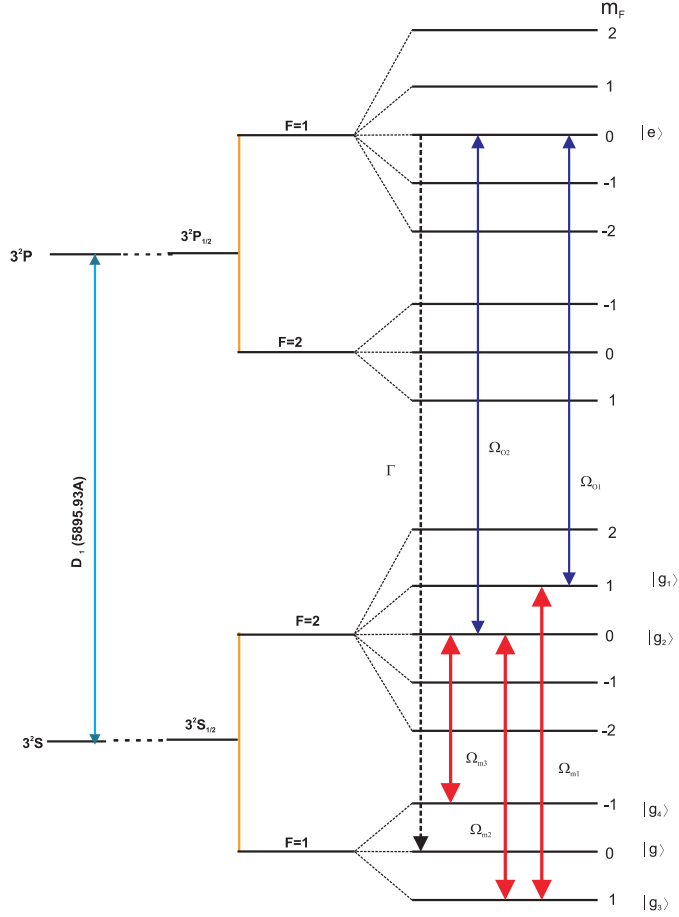}
\caption { Linkages for quintuplet spectrum of five fields and
five strongly-coupled levels of the Sodium atom. The spontaneous
emission from state $\left\vert e\right\rangle$ of the quintuplet,
shown as a dashed linkage, is responsible for the spectrum
\label{figure1}}
\end{figure}
To continue the discussion, we proceed with bare energy
statevector approach. Further, to understand and explain the
physical processes of the quantum interference with a view of the
dressing of the bare energy, the Eq.
(18) is rewritten as%
\begin{equation}
S(\Delta )\varpropto \sum_{i=1}^{5}\left\vert \frac{g^{\ast }}{\Lambda }%
\frac{[\left\vert \Omega _{B}\right\vert ^{2}-\left\vert \Omega
_{A}\right\vert ^{2}\xi _{i}^{2}+\xi _{i}^{4}]\kappa _{i}}{\Delta -\xi _{i}}%
\right\vert ^{2},  \label{30}
\end{equation}%
where $\xi _{i}$ $\left( i=1-5\right) $ are the roots of the equation%
\begin{equation}
(\Delta ^{5}-\Delta ^{3}\left\vert \Omega \right\vert ^{2}+\Delta \left\vert
\Omega _{C}\right\vert ^{4}+2\left\vert \Omega _{D}\right\vert ^{5}\cos
\varphi )^{2}+\frac{\Gamma ^{2}}{4}(\Delta ^{4}-\Delta ^{2}\left\vert \Omega
_{A}\right\vert ^{2}+\left\vert \Omega _{B}\right\vert ^{4})^{2}=0,
\label{31}
\end{equation}%
and the factor $\Lambda $ and the factors $\kappa _{i}$ $\left( i=1-5\right)
$ are given in the Appendix-B. Further, Eqs. (\ref{30},\ref{31}) are now
again simply to read. The spectrum relation contains five absolute squared
terms and twenty interference terms. The five absolute squared terms are
associated with five emission probabilities from the set of five dressed
state of the bare energy state having no dark lines in the spectrum. The
twenty interference terms are contributed by the five pathways among the
upper excited decaying dressed states responsible for generating the four
dark lines in the total spectrum of the system. We note that the
interference terms compensate the contribution from the absolute squared
terms at the locations given by the Eqs. (\ref{22},\ref{23}).

For appropriate choices of different spectroscopic variables Eqs. (\ref{32},%
\ref{33}) get the following shape%
\begin{equation}
S(\Delta )=\Gamma \sum_{i=1}^{5}\left\vert \frac{\alpha _{i}+i\beta _{i}}{%
\Delta -(\alpha _{i}^{\prime }+i\beta _{i}^{\prime })}\right\vert ^{2},
\label{32}
\end{equation}%
where in the above equation $\alpha _{i}$, $\beta _{i}$, $\alpha
_{i}^{\prime }$, and $\beta _{i}^{\prime }$ $(i=1-5)$ are the appropriate
integer values obtained due to specific choices of spectroscopic variables
of this system. Therefore, we can again readily estimate the peaks height,
their locations and their widths from the absolute squared terms. The
spectrum therefore consists of in general, five peaks\ located at $\Delta
=\alpha _{i}^{\prime }$ $(i=1-5)$ while the heights of the peaks locating at
$\alpha _{i}^{\prime }$ are given by $(\alpha _{i}^{2}+\beta
_{i}{}^{2})/\beta _{i}^{\prime 2}$ for $i=1-5$.

We again utilize the Sodium atom in its hyperfine Zeeman states of the D1
lines. Extending to six-level atomic system we select the five ground states
of D1 lines i.e., $\left\vert 3S_{1/2},F_{2}=2,m_{F}=1\right\rangle
\Leftrightarrow \left\vert g_{1}\right\rangle ,$ $\left\vert
3S_{1/2},F_{2}=2,m_{F}=0\right\rangle \Leftrightarrow \left\vert
g_{2}\right\rangle ,$ $\left\vert 3S_{1/2},F_{1}=1,m_{F}=1\right\rangle
\Leftrightarrow \left\vert g_{3}\right\rangle ,$ $\left\vert
3S_{1/2},F_{1}=2,m_{F}=-1\right\rangle \Leftrightarrow \left\vert
g_{4}\right\rangle $ and one excited state $\left\vert
3P_{1/2},F_{1}=1,m_{F}=0\right\rangle \Leftrightarrow \left\vert
e\right\rangle .$ The states $\left\vert
3S_{1/2},F_{2}=2,m_{F}=1\right\rangle \Leftrightarrow \left\vert
g_{1}\right\rangle $ and $\left\vert 3S_{1/2},F_{2}=2,m_{F}=0\right\rangle
\Leftrightarrow \left\vert g_{2}\right\rangle $ are coupled with the state $%
\left\vert 3S_{1/2},F_{1}=1,m_{F}=1\right\rangle \Leftrightarrow \left\vert
g_{3}\right\rangle $ by two microwave fields while they are coupled with the
excited decaying state $\left\vert 3P_{1/2},F_{1}=1,m_{F}=0\right\rangle
\Leftrightarrow \left\vert e\right\rangle $ by two optical fields. In
addition to this we couple another microwave field with the transition $%
\left\vert 3S_{1/2},F_{2}=2,m_{F}=0\right\rangle \Leftrightarrow \left\vert
g_{2}\right\rangle \Leftrightarrow \left\vert
3S_{1/2},F_{1}=1,m_{F}=-1\right\rangle \Leftrightarrow \left\vert
g_{4}\right\rangle $. Like the previous scheme here the linkage of the
excited state $\left\vert 3P_{1/2},F_{1}=2,m_{F}=0\right\rangle
\Leftrightarrow \left\vert e\right\rangle $ is also considered with the
ground state, $\left\vert 3S_{1/2},F_{1}=1,m_{F}=0\right\rangle
\Leftrightarrow \left\vert g\right\rangle $ via vacuum field modes. It is
again noteworthy to mention that in our scheme the order of energy levels
does not matter as we considering resonant atom fields interaction.The
nomenclature of this form the shape of a loop and a branch, another
possibility of the atom-field interaction. The same experiment of Ref. \cite%
{Zanch} is equally useful for demonstration of the results of quintuplet
spectroscopy. Here we are need to couple another microwave field with the
additional ground energy level to have five ground states. Although, the
atom-field interaction forming the configuration of a branch and a loop
however, the order of energy levels does not affect the end results for
resonant atom-field interaction of this system. Now, we need to record the
spectrum when the atom decay from the excited state to the ground.

\section{Autler-Townes Sextuplet and Higher-Ordered Multiplet Spectroscopy}

In this section we suggest the general nomenclatures for the sextuplet and
consequently for the higher-ordered multiplet Autler-Townes atomic schemes.
From the successive study of Autler-Townes spectra from the doublet to the
quintuplet we come to know that going from a lower to a next higher-ordered
spectrum we need to increase the number of atomic states by one interacting
with the required number of fields to have the required coupling states. If
there are N coupled states then there is the generation of N dressed states
associated with the decaying bare state of the scheme under investigation
regardless of the number of fields. Generally this procedure is necessary to
increase the number of dressed states by one. In this way we may be able to
split the decaying bare energy state to any number of dressed energy state
depending on the number of energy levels of the atomic system, the nature of
the required fields and the nature of their interaction. Earlier, in our
previously proposed schemes for the quartuplet and quintuplet spectroscopy
we based our schemes on D1 line of the hyperfine-structured Sodium atom to
have simple probability loss. Due to this plausible simplification, the
atomic systems are the ideal one having one decaying levels and are handled
analytically by a statevector method even for these complicated systems. It
worthy to mention that these simplified systems are necessary for displaying
the ideal behavior. The potentiality of these systems is unquestionable
because the multiple uncontrollable decaying processes may lead to
complications to disturb the ideal behavior of the involved dynamics in the
systems under investigation. The states of these simple schemes correspond
to pure states and the obtained results are mathematically very easy to
explain various aspects of the phenomenon of spontaneous emission processes
in the higher-ordered multiplet spectroscopy.

To go beyond and to suggest schemes of higher-ordered multiplet spectrum,
for example the sextuplet atomic scheme, we need six ground hyperfine energy
states and an excited energy state of the sodium atom. Luckily this type of
nomenclature can also be found in the sodium D1 lines in hyperfine Zeeman
splitting. Six ground states can be selected from the eight closely spaced
Zeeman levels associated with the lower level of the fine energy level of
the sodium D1 line and one from the allowed ones of the excited states. This
seven energy states scheme can be driven by two optical and five microwave
fields either in a loop or a loop plus a branch structured configuration
using the real sodium atom. This then forms a scheme of having simple
probability loss and can be handled by the previously presented approach to
have ideal behavior. Due to this atom-field interaction, the only one
decaying bare energy states will split into six dressed energy states which
will then interfere destructively among themselves to create five dark lines
in the spectrum. These dark lines will then divide the spectrum into six
components.

Furthermore, if the fine structured sodium atom is considered, for example
the ones of Zuo et al. which requires a loop plus a branch structured
involving three decay rates such that the system is strictly subjected to
the resonant atom-field interaction with no requirement of microwave fields.
The configuration then forms the shape of folded NV shaped atomic system.
This scheme involves three decay processes and corresponds to a mix state
and may not be handled by the previously presented statevector approach.

The system becomes more complicated due to the next increase in the number
of coupled atomic energy levels and the increase in the number of the driven
laser fields for higher ordered multiplet spectroscopy. In this situation it
will not be very easy to avoid decaying processes and the other processes of
decoherence during an experiment and may affect the ideal behavior \cite%
{Zhu-Narducci,Fazal}. However, if these complications are very large in its
rates then the decoherence processes may completely eliminate the dark lines
feature from the spectra. The process of quantum destructive interference
among the dressed energy states will then be completely washed out from the
system. It will also become very hard to handle the problems with the
statevector methods and may be dealt with the density matrix formalism or
using generalized Bloch equations of motion for the system under
investigation using numerical simulation. This idea is extendable into any
higher-ordered multiplet spectrum. Proposing such a multiplet scheme
requires careful selection of the laser fields, number of atomic energy
levels of the atom or molecule and their appropriate interaction such that
there is no violation of parity except the ones that nature permit.

\begin{figure}[t]
\centering
\includegraphics[width=3.5in]{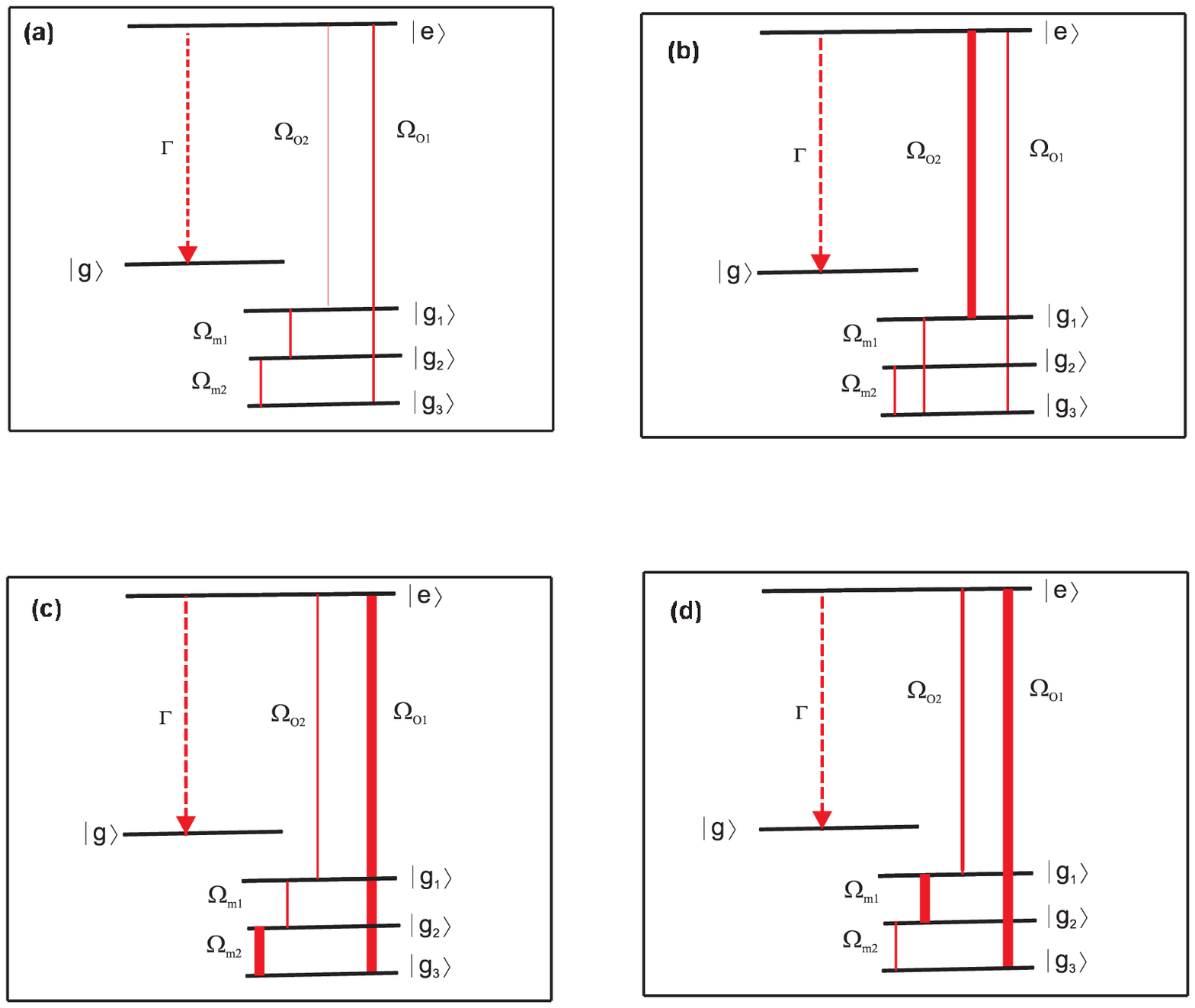}
\caption { Linkage patterns and relative strengths of four driving
fields for quartet spectra of Fig. 4. \textbf{(a)} all Rabi
frequencies are equal to $0.5\Gamma $. \textbf{(b) }The single
Rabi frequency $\left\vert \Omega _{o_{2}}\right\vert $ is
$4\times $ the others$.$ \textbf{(c)} Rabi frequencies $\left\vert
\Omega _{o_{1}}\right\vert $ and $\left\vert \Omega
_{m_{1}}\right\vert $ are $4\times $ the others$.$ \textbf{(d)
}Rabi frequencies $\left\vert \Omega _{o_{1}}\right\vert $ andr
$\left\vert \Omega _{m_{2}}\right\vert $ are $4\times $ the
others$.$ In these linkage plots the vertical positions of the
lines associated with energy states is irrelevant, unlike Fig. 1
where vertical position is associated with energy.
\label{figure1}}
\end{figure}

\section{Discussion}

In this section we discuss our main results presented in the text. In
presenting these results we take support of the analytical results and the
numerical estimations regarding various facts of Autler-Townes multiplets
spectra. We specially underlined the techniques in the text that how the
widths of all the spectral components (bright lines) in all the schemes can
be estimated from the numerical analysis. We preferred this way of analysis
as due to complex nature of atom-field interaction, the analytical forms of
the widths of all spectral components in their corresponding schemes are
very hard to calculate and are not presentable.

\subsection{The Quartuplet Spectrum}

First of all we inspect the generality of quartuplet atomic system. The
analytical result obtained for this system is given by Eq. (\ref{5}). The
atomic system reduces to a two-level atom interacting with the vacuum field
modes when all the four laser fields are vanished. This results in a
Lorentzian line-shape spectrum whose height is $4\left\vert g_{\mathbf{k}%
}\right\vert ^{2}/\Gamma ^{2}$ and the FWHM is $\Gamma $ \cite{Book1}.
Furthermore, the result reduces to the form

\begin{equation}
S\left( \Delta \right) =\frac{\left\vert g\right\vert ^{2}\Delta ^{2}}{%
\left( \Delta ^{2}-\left\vert \Omega _{o_{1}}\right\vert ^{2}\right) ^{2}+%
\frac{\Gamma ^{2}}{4}\Delta ^{2}}.  \label{33}
\end{equation}%
when only $\left\vert \Omega _{o_{2}}\right\vert $, $\left\vert \Omega
_{m_{1}}\right\vert $ and $\left\vert \Omega _{m_{2}}\right\vert $ are set
to zero. In this case the atomic system reduces to a three-level atom
interacting with a laser field and results in the well known Autler-Townes
doublet spectrum \cite{Zhu-Narducci} showing its generality. Furthermore,
this scheme cannot be reduced to the results of Autler-Townes triplet
spectrum as all the atomic systems are subjected to no parity violation
condition except the ones that nature permit. Therefore, there is symmetry
among the even numbered Autler-Townes multiplet schemes when and only when
the schemes are based on fine structured atomic system while avoiding the
parity violations. The higher even numbered multiplet schemes then reduce to
the analytical results of the preceding even numbered multiplet schemes. For
example in the present case the results of the quartuplet spectrum is
reduced to the results of doublet scheme as shown above. However, this
symmetry is valid only for spectrum beyond Autler-Townes triplet in the
present case.

\begin{figure}[t]
\centering
\includegraphics[width=3.5in]{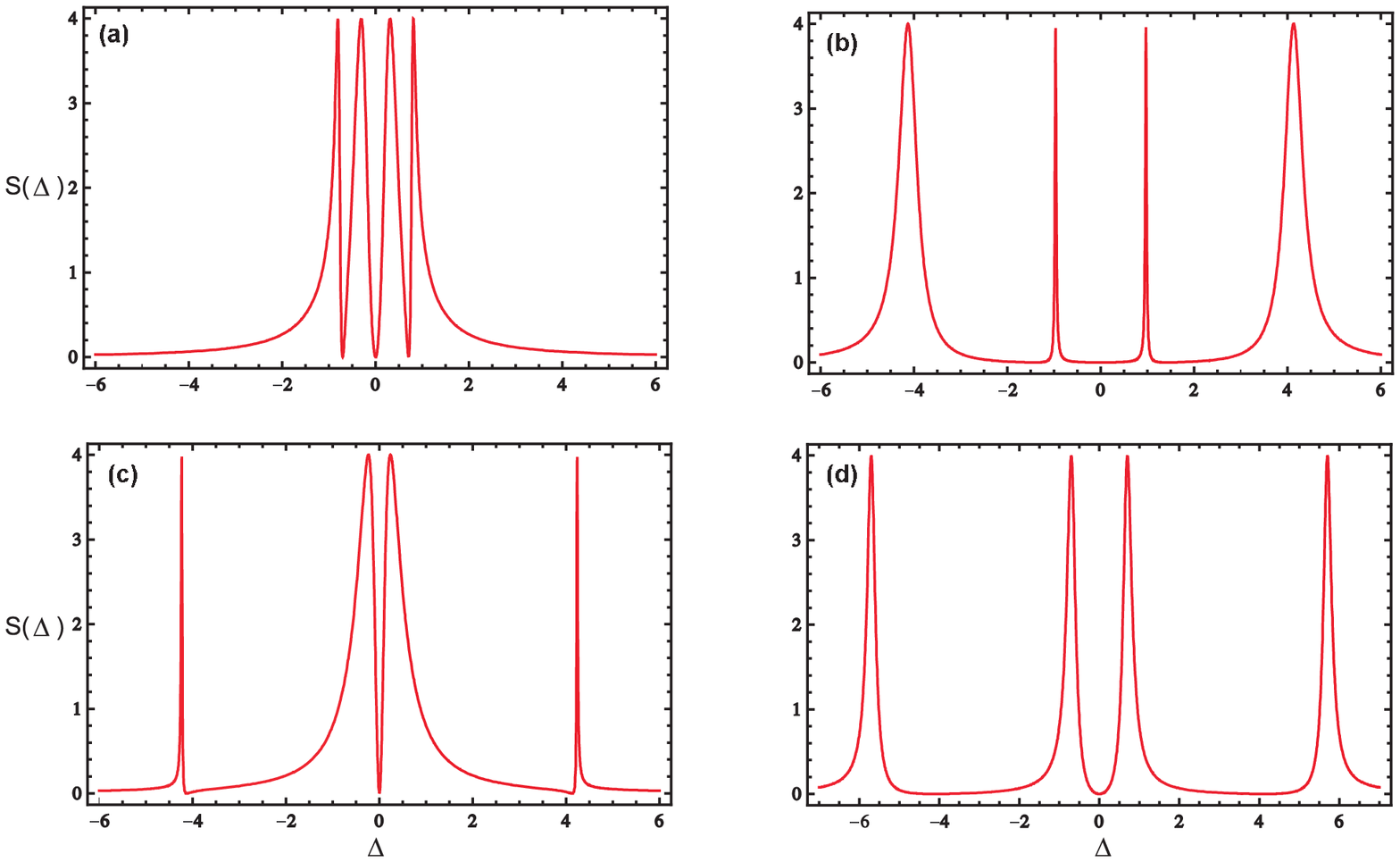}
\caption { Quartuplet\textbf{\ }spectrum for four fields and five
states for the linkages of Fig. 3 with phase $\varphi =2\pi $ and \textbf{(a)%
} $\left\vert \Omega _{o_{1}}\right\vert =\left\vert \Omega
_{o_{2}}\right\vert =\left\vert \Omega _{m_{1}}\right\vert
=\left\vert \Omega _{m_{2}}\right\vert =0.5\Gamma $. \textbf{(b)}
$\left\vert \Omega _{o_{1}}\right\vert =\left\vert \Omega
_{m_{1}}\right\vert =\left\vert \Omega _{m_{2}}\right\vert
=1\Gamma $ and $\left\vert \Omega _{o_{2}}\right\vert =4\Gamma $.
\textbf{(c)} $\left\vert \Omega
_{o_{1}}\right\vert =\left\vert \Omega _{m_{1}}\right\vert =4\Gamma $ and $%
\left\vert \Omega _{o_{2}}\right\vert =\left\vert \Omega
_{m_{2}}\right\vert =1\Gamma $. \textbf{(d)} $\left\vert \Omega
_{o_{1}}\right\vert =\left\vert \Omega _{m_{2}}\right\vert
=1\Gamma $ and $\left\vert \Omega _{o_{2}}\right\vert =\left\vert
\Omega _{m_{1}}\right\vert =4\Gamma $. \label{figure1}}
\end{figure}

Now we proceed with the discussion of the results for the
quartuplet spectrum. In Fig. (4), we present the linkages of the
driving fields with the atomic system according to our proposal
presented in section 3. However, in these linkages plots the
vertical positions of the lines associated with actual position in
the sodium hyperfine structured atomic system is irrelevant. Also,
in this quartuplet scheme, the linkages of the coherently strong
driving fields associated with their respective Rabi frequencies
are shown by the thick lines while the thin are shown for the weak
fields under consideration. These are displayed for the better
understanding of the physics of the spectrum for this quartuplet
as the order of energy levels in the resonant atom-field
interactions does not matter. We note that the three dark lines
are always there regardless of the strengths of the four driving
fields. The four spectral components can obviously be seen
separated by the three dark lines at the locations $\Delta $\ =
$0,$ $\pm \sqrt{\left\vert \Omega _{m_{1}}\right\vert
^{2}+\left\vert \Omega _{m_{2}}\right\vert ^{2}}$ for the small
values of the Rabi frequencies as shown in Fig. (5a). In Fig. (4a)
the four equally linked weak fields with the atomic transition
frequencies are shown by the thin lines. The spectral components
are equally spaced on the frequency axis having small separation.
These three dark lines appear due to the interference effect among
the four decaying dressed states created by the atom-field
interaction. One dark line is always located at the central line.
However, the other two which lying symmetrically on the positive
and negative frequency axes of the spectrum are effectively
controlled by the variation of $\left\vert \Omega _{m_{1}}\right\vert $ and $%
\left\vert \Omega _{m_{2}}\right\vert $. This symmetry arises due to the
resonance coupling of the driving fields with their respective atomic
transitions frequencies. However, if the fields are detuned from these
transitions then the spectrum will become asymmetric. Here, we limit our
discussion only to the resonance cases due their various advantages.
Further, both the dark lines and the effective Rabi frequency for this
system are independent of the Rabi-frequency $\left\vert \Omega
_{o_{1}}\right\vert $ which means that the quartuplet nature of the spectrum
is not altering if $\left\vert \Omega _{o_{1}}\right\vert =0$. However, the
phase dependent term, $\left\vert \Omega _{o_{1}}\right\vert \left\vert
\Omega _{o_{2}}\right\vert \left\vert \Omega _{m_{1}}\right\vert \left\vert
\Omega _{m_{2}}\right\vert \sin \varphi $ appearing at the denominator of
the spectrum Eq. (\ref{5}) is vanished eliminating the phase effect from the
system. Therefore the loop structure of the atom-field interaction is
important for the preserving the phase effect in the system, a parameter
which may have numerous advantages for future studies related to this work
\cite{Fazal111}. Furthermore, using Eq. (\ref{5}) we note that the FWHM of
these spectral components for the same strength of the driving laser fields
are always the same and are equal to one-fourth of the Einstein's decay rate
$\Gamma .$ The sum of these widths is equal to $\Gamma $ which is the
standard value as obtained by Weisskopf-Wigner theory for the spontaneous
emission of two-level decaying atom due to vacuum. However, for stronger but
same driving fields, the dark lines feature of the spectrum equally widens,
with the expected behavior of the upper excited four dressed states, whose
spacing are controlled by the Rabi frequencies of all the four driving
fields. Moreover, the FWHM of each of the four spectral components in this
case is also $\frac{\Gamma }{4}$ and obeying the Weisskopf-Wigner theory for
its total width. Obviously, the shapes of the individual spectral component
in these cases and in the coming up are not Lorentzian. They are the
examples of Breit-Wigner or Fano profiles. It is now obvious to describe
that when all the Rabi frequencies are equal then the widths of all the
peaks are then all equal and they are separated equally by greater amounts.
Consequently the peaks are more widely spaced while their widths are
unaffected. Generally, in the quartuplet atomic scheme there are always four
spectral lines separated by the dark lines and associated with the four
dressed states created by the interaction of four fields with the atom.
However, it is worthwhile to mention that the number of the dressed states
and the number of the fields which are equal in this system, is accidental.
In principle, one dressed state can be associated with one strongly linked
state by the fields regardless of its numbers. If there are N linked atomic
states with the driving fields then there must be N dressed states. For
example, the very famous scheme for spontaneous emission cancellation \cite%
{Zhu-Scully,Xia,Mart} involves only one field in a four level atomic system.
However, the strong linkage of this single field with three atomic states
creates three dressed states while there is only one driving field in the
system. Another very famous scheme is the Autler-Townes doublet where single
strong coherent driving field linked with the two bare states and thus
generates two dressed states for the one decaying bare state.

\begin{figure}[t]
\centering
\includegraphics[width=4.5in]{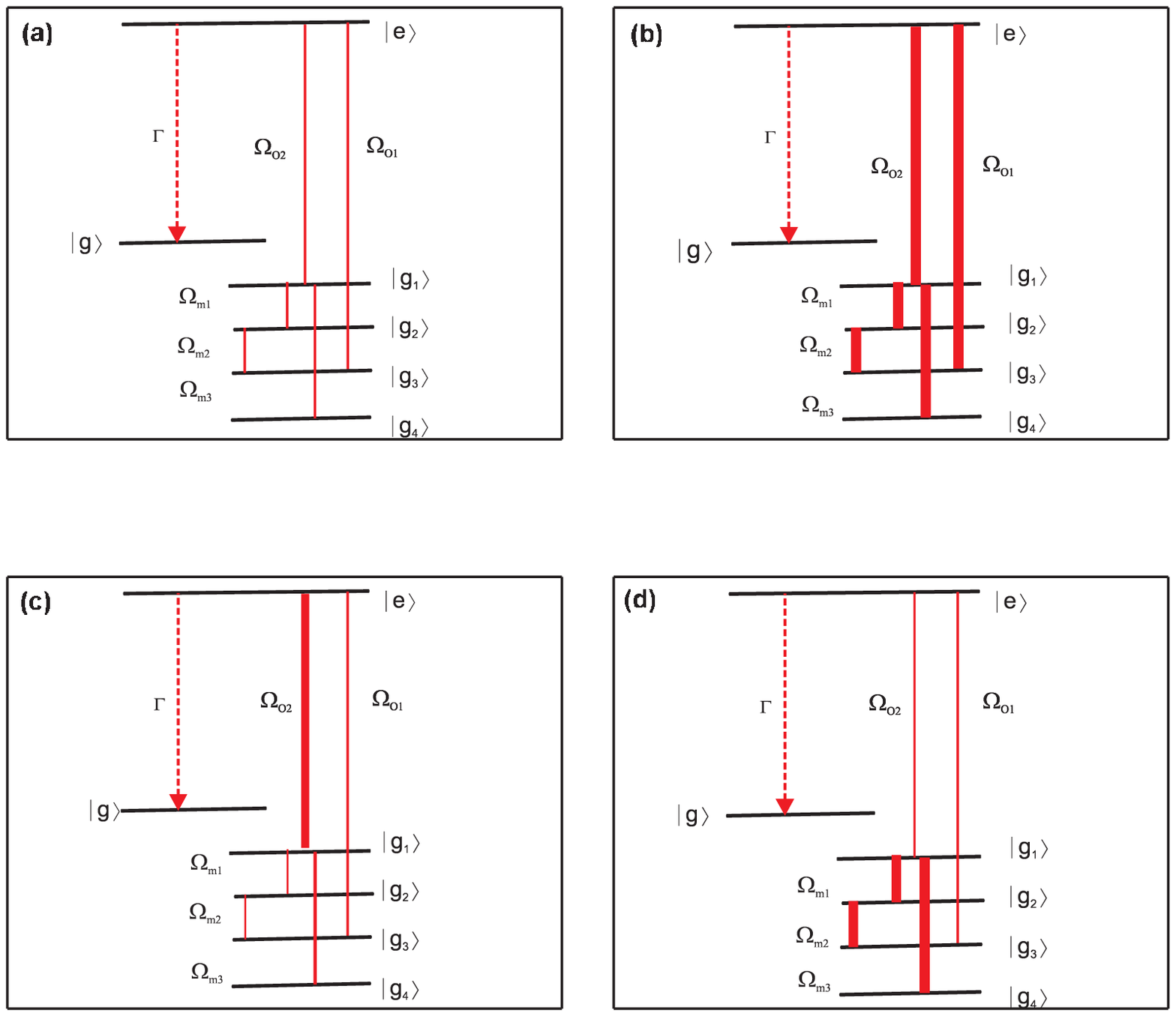}
\caption { Linkage patterns and relative strengths of five driving
fields for quintuplet spectra of Fig. 6. \textbf{(a)} all Rabi
frequencies are equal to $0.5\Gamma $. \textbf{(b)} Rabi
frequencies are strong, equal to $3\Gamma .$ \textbf{(c)} Rabi
frequencies $\left\vert \Omega _{o_{2}}\right\vert $ is $4\times $
all the others$.$ \textbf{(d)} Rabi frequencies $\left\vert \Omega
_{m_{3}}\right\vert $, $\left\vert \Omega
_{m_{2}}\right\vert $ and $\left\vert \Omega _{m_{1}}\right\vert $ are $%
4\times $ the others. In these linkage plots the vertical
positions of the lines associated with energy states is
irrelevant, unlike Fig. 2 where vertical position is associated
with energy \label{figure1}}
\end{figure}

Next, we explore the nature of the line-width narrowing and broadening in
different spectral components of the emission spectra of the quartuplet
scheme. We explain clearly the coherence effect of the driving fields and
its distinction regarding the opposite effects of the line-width narrowing
and broadening associated with their appropriate driven fields respectively.
We note the extreme line width narrowing in all the four peaks of the
spectrum with the change in relative strengths of the coherently driving
fields. The detailed numerical analysis of the spectrum Eq. (\ref{11}) for
the different sets of parameters of the system variable leads us to a
similar response as compared with the analytical results. The FWHM of the
central two peaks associated with the two central dressed states of the set
of four dressed states for the decaying bare energy state start to decrease
with the increase of either $\left\vert \Omega _{o_{1}}\right\vert $ or $%
\left\vert \Omega _{0_{2}}\right\vert $. They become lesser and
lesser with the more increase of the corresponding Rabi frequency
as shown by the Fig. 5(b). The linkages of one strong field and
three weak fields with atomic system are displayed in Fig. 4(b) by
a thick line and three thin lines, respectively. This effect of
spectral narrowing becomes double for both the central peaks when
there is simultaneous increase in both the Rabi frequencies.
However, the widths of the side peaks of the set of four dressed
states become larger and larger accordingly. Furthermore, we also
noticed that the decrease in widths of the two central peaks
always compensates the increase in the widths of the side two
peaks of the spectrum for both the individual or simultaneous
increase of the Rabi frequencies. This is in accord with the
theory of Weisskopf-Wigner and the sum of these
widths is always equal to Einstein's decay rate. The plot of the Eq. (\ref{5}%
) manifests the behavior what we predicted from the numerical
analysis of the widths of the four spectral components. The four
linkages for the two weak fields and the two strong fields are
also shown by the Fig. 4(c). Further, the widths of the two
central peaks are always the same and these decreases to the same
value of $0.05\Gamma $ for an optimum value of the Rabi frequency
$\left\vert \Omega _{o_{1}}\right\vert $, while the widths of
the side two peaks increase in the same proportion and is equal to $%
0.45\Gamma $ for each one respectively. The same effects are seen when $%
\left\vert \Omega _{o_{2}}\right\vert $ is increased instead of
$\left\vert \Omega _{o_{1}}\right\vert .$ Moreover by increasing
both the Rabi frequencies, the narrowing effect is doubled with
the side peaks displaced more from the central one [see Fig.
5(c)]. While increasing the Rabi frequencies $\left\vert \Omega
_{m_{1}}\right\vert $, $(\left\vert \Omega _{m_{2}}\right\vert )$
or both, the FWHM of the two central peaks of the set of four
peaks of the spectrum starts to increase drastically with a double
effect in the simultaneous case. However, it reduces the widths of
the two side peaks correspondingly. As a result we obtain extreme
line-width narrowing in the side two peaks and extreme line-width
broadening in the two central peaks for optimum value of these
Rabi frequencies. This increase and decrease in the two central
and the two sides peaks are again always in the same proportion
and obey Weisskopf-Wigner theory. The maximum possible increase in
the widths of the two central peaks in this case is $0.45\Gamma $
for each one of the two, whereas for the sides peaks it decreases to $%
0.05\Gamma $ for each one of the two. However, when the linkages
pattern of the driving fields like the one shown in the atomic
scheme of Fig. 3(d), the spectrum displays irregular behavior
regarding the widths of the four components of the spectrum and
the position of the dark lines. The reason is obvious. The one
stronger field having narrowing effect in one component compensate
the broadening effect of the other equally stronger field on the
same component see Fig. 5(d). However, the peaks take the position
according to the strengths of the fields. The fields which are
directly linked with the decaying level play the role of
displacing away (closing together) the two central peaks while the
fields which are not directly linked with the decaying level play
the role in displacing away (closing together) the sides peaks of
the spectrum when fields strength are increasing (decreasing).
Physically, this is reasonable as the splitting of the dressed
states are directly controlled by the strengths of the driving
fields but in a symmetrical way. In fact, this is a relative
phenomenon for this system regarding the effect of relative
strengths on the spacing of the dressed states for this system.
However, the combinational effect can then be dealt accordingly.
This makes sense to having unequal spacing with the irregular
strength of the driving fields avoiding the principle stated
above. Further, it is also worthwhile to note that the numerical
analysis of the widths of different spectral components is
sufficient to explore the physics of spectral narrowing and
broadening arises due to the quantum interference mechanism in the
atomic system. However, it is not difficult to plot the response
of widths with the different Rabi frequencies of the system by
following the method discussed in Ref. \cite{Fazal00}.

\subsection{The Quintuplet Spectrum}

The analytical results obtained for the quintuplet atomic system are given
by the Eq. (\ref{16}). This atomic system also reduces to a two level atom
interacting with the vacuum fields when all the five laser fields are
vanished. This results in a Lorentzian line-shape spectrum whose height is $%
4\left\vert g_{\mathbf{k}}\right\vert ^{2}/\Gamma ^{2}$ and the FWHM is $%
\Gamma $ \cite{Book1}. In this case too, the result reduces to the form of
Autler-Townes doublet when the Rabi frequencies $\left\vert \Omega
_{m_{1}}\right\vert $, $\left\vert \Omega _{m_{2}}\right\vert $, $\left\vert
\Omega _{m_{3}}\right\vert $ and $\left\vert \Omega _{o_{2}}\right\vert $ in
Eq. (\ref{16}) become zero. Further when the Rabi frequencies $\left\vert
\Omega _{o_{2}}\right\vert $ and $\left\vert \Omega _{m_{3}}\right\vert $ in
the analytical expression of this system are set to zero then the
spontaneous emission spectrum reduces to the expression of Autler-Townes
triplet spectrum \cite{Fazal,Fazal00} given by

\begin{equation}
S\left( \Delta \right) \varpropto \frac{\left\vert g\right\vert ^{2}\left(
\left\vert \Omega _{m_{1}}\right\vert ^{2}-\Delta ^{2}\right) ^{2}}{\left(
\Delta ^{3}-\left( \left\vert \Omega _{o_{1}}\right\vert ^{2}+\left\vert
\Omega _{m_{1}}\right\vert ^{2}+\left\vert \Omega _{m_{1}}\right\vert
^{2}\left\vert \Omega _{m_{2}}\right\vert ^{2}\right) \Delta \right) ^{2}+%
\frac{\Gamma ^{2}}{4}\left( \left\vert \Omega _{m_{1}}\right\vert
^{2}-\Delta ^{2}\right) ^{2}}.  \label{34}
\end{equation}%
Furthermore, the quintuplet scheme cannot be reduced to the results of
Autler-Townes quartuplet spectrum as all the systems are independent of
parity violation. Therefore, there is symmetry among the odd numbered
Autler-Townes multiplet schemes when and only when the schemes are based on
the hyperfine structured atomic system having no parity violations. The
analytical results for higher odd numbered schemes will reduce to the
analytical results of the preceding odd numbered Autler-Townes multiplet
schemes. For example in the present case the result of the quintuplet
spectrum is reduced to the result of triplet scheme as shown above. However,
this symmetry is valid only for spectrum beyond Autler-Townes quartuplet
spectrum in the case of odd numbered Autler-Townes multiplet schemes.

\begin{figure}[t]
\centering
\includegraphics[width=4.5in]{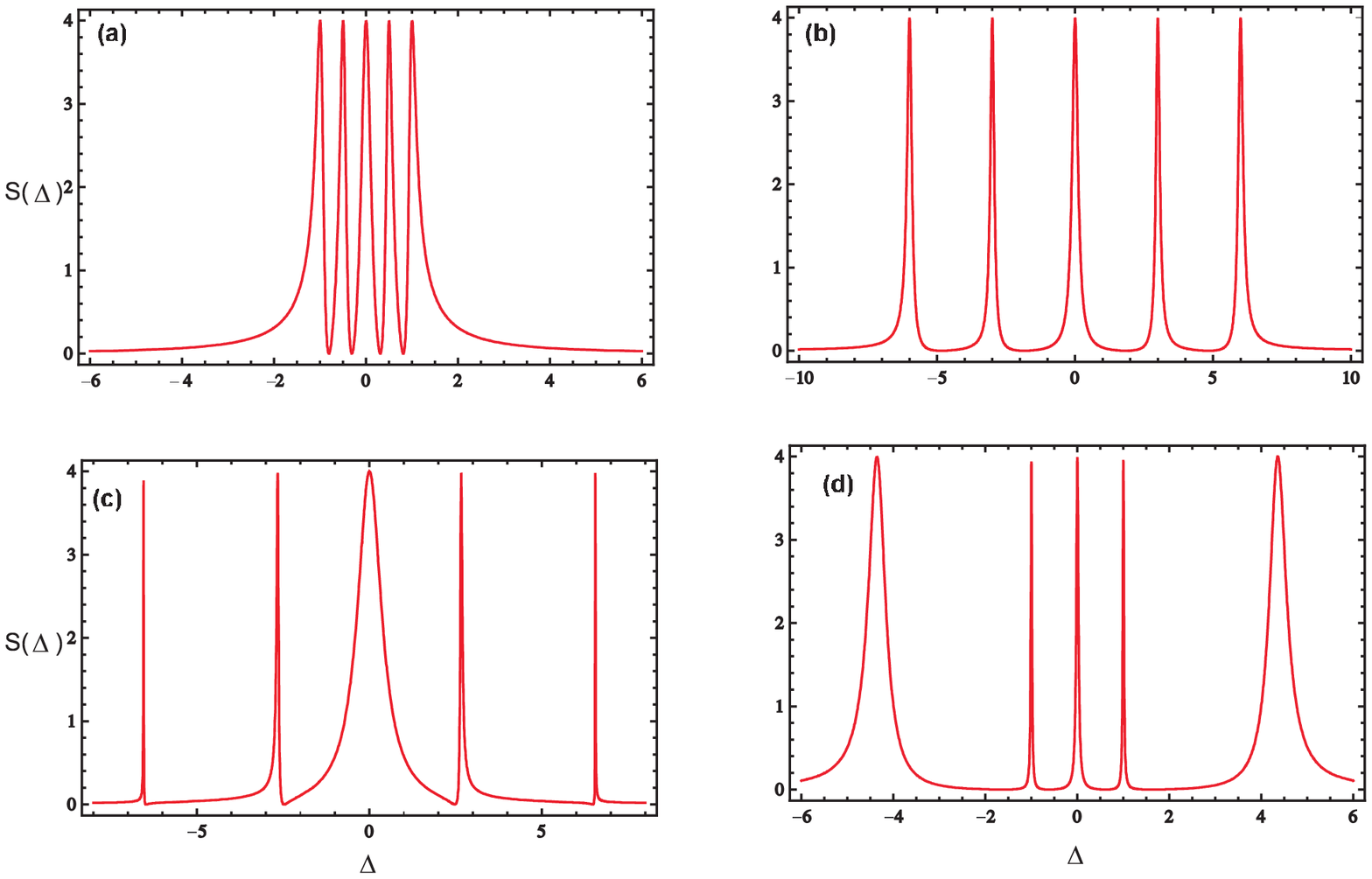}
\caption { LQuaintuplet\textbf{\ }spectrum for four fields and
five states for the linkages of Fig. 5 with phase $\varphi =2\pi $
and \textbf{(a)} $\left\vert \Omega _{o_{1}}\right\vert
=\left\vert \Omega _{o_{2}}\right\vert =\left\vert \Omega
_{m_{1}}\right\vert =\left\vert \Omega _{m_{2}}\right\vert
=\left\vert \Omega _{m_{3}}\right\vert =0.5\Gamma $. \textbf{(b)}
$\left\vert \Omega _{o_{1}}\right\vert =\left\vert \Omega
_{o_{2}}\right\vert =\left\vert \Omega _{m_{1}}\right\vert
=\left\vert
\Omega _{m_{2}}\right\vert =\left\vert \Omega _{m_{3}}\right\vert =3\Gamma $%
. \textbf{(c)} $\left\vert \Omega _{o_{1}}\right\vert =4\Gamma $ and $%
\left\vert \Omega _{o_{2}}\right\vert =\left\vert \Omega
_{m_{1}}\right\vert =\left\vert \Omega _{m_{2}}\right\vert
=\left\vert \Omega _{m_{3}}\right\vert =1\Gamma $. \textbf{(d)}
$\left\vert \Omega
_{o_{1}}\right\vert =\left\vert \Omega _{o_{3}}\right\vert =1\Gamma $ and $%
\left\vert \Omega _{m_{3}}\right\vert =\left\vert \Omega
_{m_{2}}\right\vert =\left\vert \Omega _{m_{3}}\right\vert
=4\Gamma $. \label{figure1}}
\end{figure}

In Fig. (5), we present the linkages of the driving fields with
the atomic system proposed for this system. In these Linkages
plots the vertical positions of the six lines associated with
actual positions in the sodium hyperfine structured atom (see Fig.
(3)) are irrelevant. In this quintuplet scheme, the linkages of
the coherently strong driving fields associated with their
respective Rabi frequencies are shown by the thick lines while
they are thin for the weak fields. Here, we observed four dark
lines in the spectrum irrespective of the strength of the five
driving fields. The
locations of these four dark lines can be inspected very easily from Eqs. (%
\ref{22},\ref{23}) which split the spectrum into five spectral
components as shown by the Fig. 7 (a). Their relevant linkages
pattern for all the weak fields is shown in Fig. 7(a). The source
of these dark lines is the interference mechanism among the five
paths from the five dressed decaying states created by the
atom-field interaction. Again, all the dark lines as
well as the effective Rabi frequency are independent of Rabi-frequency $%
\left\vert \Omega _{o_{2}}\right\vert $ and means that the quintuplet nature
of the spectrum is not altering if $\left\vert \Omega _{o_{2}}\right\vert =0$%
. However, the phase dependent term, $\left\vert \Omega _{o_{1}}\right\vert
\left\vert \Omega _{m_{1}}\right\vert \left\vert \Omega _{m_{2}}\right\vert
\left\vert \Omega _{m_{3}}\right\vert \left\vert \Omega _{o_{2}}\right\vert
\cos \varphi $ appearing at the denominator of the spectrum Eq. (\ref{18})
is vanished, washing out the phase effect from the system. Further, the
naturally conversion of $\sin \varphi $ to $\cos \varphi $ in this system is
satisfactory as the increase in the energy level of the system changes the
origin of the phase by $\frac{\pi }{2}.$ This change in the phase arises due
to the term in the calculations of the spectrum for the quintuplet scheme
appearing as a complex conjugate when compared with its related term in the
quartuplet scheme. The FWHM of each of the five spectral components for the
same strength of the driving laser field is always the same and is equal to $%
\frac{\Gamma }{5}.$ The sum of these five widths is equal to
$\Gamma $ which agrees with Weisskopf-Wigner theory. However, for
stronger but same driving fields, the dark line features of the
spectrum widens equally, with the expected behavior of the upper
excited five dressed states, whose spacings are controlled by the
Rabi frequencies of all the five driving fields. This is shown in
Figs. 7(b) where the linkage of the strong field display as thick
line while for the weak fields these are thin see Fig. 6(b).
Moreover, the FWHM of each of these five spectral components in this case is also $%
\frac{\Gamma }{5}$ and obeying the Weisskopf-Wigner theory for its
total width. Again, we explore the nature of the line-width
narrowing and broadening in different spectral components of the
emission spectra of the quintuplet scheme. The response of the
quintuplet spectrum regarding the line-width narrowing and
broadening of the five spectral components with the relative
strength of the driving fields are also enormous. Using Eq.
(\ref{34}), we estimated numerically that by increasing the Rabi
frequency $\left\vert
\Omega _{m_{1}}\right\vert $, $\left\vert \Omega _{m_{2}}\right\vert $ and $%
\left\vert \Omega _{m_{3}}\right\vert $ the FWHM of the central
peak of the spectrum starts to increase drastically. All these
three Rabi frequencies individually have the same effect on the
width of the central peak. However, their effects are adding if
the increase in these Rabi frequencies are considered
simultaneously in two or three combination while these displace
the side peaks more from the central one. Correspondingly, these
reduce the widths of the four side peaks to obey Weisskopf-Wigner
theory. As a result we obtained an extreme line-width narrowing in
the sides four peaks and an extreme line-width broadening in the
central peak for optimum value of the Rabi frequency as shown by
Fig. (7c). The linkage of the fields with atomic system is also
shown in Fig. (6c). The maximum possible increase in the width of
the central peak is $0.7\Gamma $, whereas for the side peaks the
widths decrease to $0.1\Gamma $ and $0.05\Gamma $ respectively. We
next note, the FWHM of the central three peaks associated with the
three central dressed states of the set of five dressed states of
the decaying bare energy state start to decrease with the increase
of $\left\vert \Omega _{o_{1}}\right\vert $ and becomes lesser and
lesser with more increase of the Rabi frequency. However, it
becomes larger and larger for the sides two peaks of the set of
five peaks. Furthermore, we also noticed that the decrease in
widths of the three central peaks always compensates the increase
in the widths of the side two peaks of the set of five peaks of
the
spectrum obeying the theory of Weisskopf-Wigner. The plot of the Eq. (\ref%
{18}) manifests the behavior of what we predict from the analysis
of the widths of the five spectral components. The widths of the
three central peaks are always the same and these decreases to a
value of $0.067\Gamma $ for an optimum value of the Rabi frequency
$\left\vert \Omega _{o_{1}}\right\vert .$ However, the widths of
the side two peaks increase in the same proportion and is equal to
$0.4\Gamma $ for each ones. We also provide the linkages pattern
of the driving fields regarding their strengths in the Fig. 6(d).
The same effect is seen when $\left\vert \Omega _{o_{2}}\right\vert $ is increased instead of $%
\left\vert \Omega _{o_{1}}\right\vert .$ Moreover by increasing
both the Rabi frequencies, we have the doubled effect regarding
the spectral narrowing. However, the side peaks are more displaced
from the central one in this case as shown in Fig. 7(d).

In the discussion of this and the previous system we consider the simplest
systems which display ideal behavior obeying Weisskopf-Wigner theory. These
then correspond to pure states and are handled via a statevector method.
However, if mixed states are considered where the decay processes from the
multi levels are involved, the Weisskopf-Wigner theory is no longer obeyed.
This is due to the fact that the additional contributions of the losses at
high rates, are added to the systems in the form of increase of the widths
of the spectral components. Also this contribution minimizes the effect of
the interference mechanisms in the system. However, these behavior does not
correspond to our quartuplet and quintuplet systems as they have simply
probability losses. Generally, this is natural as all the ideal behavior are
always disturbed by any kind of decoherence processes in any system.
However, in these presented systems the broadening in the spectral
components are the possibly minimum one allowed by Weisskopf-Wigner theory
for the vacuum fields (not modified fields) fluctuations only

\section{\textbf{Comparison}}

Obviously, the response of the widths of the five spectral components of the
spectrum for the quintuplet scheme is different from the four components of
the spectrum for the quartuplet scheme with the relative strengths of their
corresponding driving fields. The detailed study of the multiplet
spectroscopy reveals that there is always decrease in the widths of some
components with the increase in the Rabi frequencies associated with those
driving fields which are coupled directly with the decaying energy level of
the system under investigation. The driving fields which are not coupled
directly with the appropriate decaying energy level of the system have the
opposite effects on the same components of the spectrum. For example, in
triplet spectroscopy where only three driving fields are involved in which
the two directly coupled fields with the decaying energy level have a
similar effect on the central peak of the triplet while there is only one
field which is indirectly coupled with the decaying energy level of the
system and have the opposite effect on the same spectral components \cite%
{Fazal,Fazal00}. 
In quartuplet spectrum, there are again two directly coupled fields with the
decaying energy level while the indirectly coupled fields are also two but
have the opposite effect on the same spectral components of the system.
Similarly in the quintuplet spectrum, there are again two directly coupled
fields with the decaying state and have similar narrowing effect on a
spectral component. However, here the indirectly coupled fields are three
having the broadening effect on the same spectral component of the system.
This procedure is valid for any ordered multiplet scheme, where the effect
of two directly coupled fields with a decaying energy level will always have
the opposite effect on a spectral component as compared with the rest of the
fields indirectly coupled with the decaying energy state of the system under
the investigation. Generalizing, we can conclude that the same principle can
be applied to a system having multiple decaying processes provided the one
of the all decays is dominant over the others. Physically we can understand
this behavior from the principle that the coherence of one field induces the
others following some symmetric way concerning with nature of the coupled
fields with the atom. Thus the effect of the combination of all the fields
of a system narrows some spectral components while broadens the remaining to
obey Weisskopf-Wigner theory. This phenomenon is completely different from
the spontaneously generated coherence, a phenomenon arising due to coherence
in the incoherent decay processes.

The effects of the relative strength of the driving fields on the position
of different spectral components in Autler-Townes multiplet spectroscopy is
also very important to discuss. Here we outline the general role of how the
position of spectral components are affected with the relative strength of
the driving fields. The role of the driving fields plays in a systematic
manner on the different spectral components of the spectrum for a given
atomic system depending on the strengths of the fields linked. Obviously the
positions of the spectral components are traced back to the spacing among
the different dressed states of a bare state created by the interactions of
these fields with the atomic system either in the form of a loop or branch
plus loop. The role of the different driving fields in the multiplet spectra
is more complex than the field in the Autler-Townes doublet. However, due to
fields generated coherence we can easily observe the correlation of the
peaks position in the Autler-Townes multiplet spectra with the relative
strength of the coupled laser fields . The reason is obvious. For example,
the fields which are directly linked with the decaying level play the role
of displacing away (displacing together) the two central peaks of the set of
four peaks while the fields which are not directly linked with the decaying
level play the role of displacing away (displacing together) the sides peaks
of the spectrum with the increase (decrease) of the fields strength.
Physically, this is reasonable as the splitting of the dressed states are
directly controlled by the strength of the driving fields. In fact, this is
a relative phenomenon for this system regarding the effect of relative
strengths on the spacing of the dressed states for this system. However, the
combinational effect can then be dealt accordingly. This makes sense to
having unequal spacing with the irregular strength of the driving fields
avoiding the principle stated above. Furthermore, if these correlations in
the relative strength are carried in a systematic manner then the increase
or decrease in the spacing of the peaks positions are also in a systematic
manner following the principle stated above. If there are odd number of
spectral components, for example the quintuplet which has one spectral
component always at line center while the rest follow the same principle
stated for the quartuplet. However, it is not difficult to generalize this
principle for locations of spectral components of any order multiplet
spectroscopy provided the system is subjected to the fine structured
configuration of naturally existing atoms of simple probability loss.

\subsection{Autler-Townes Multiplet Spectra and The Existing Related Works}

Although there are symmetries among the different atomic schemes presented
for different Autler-Townes multiplet spectra in this paper. However, these
schemes are not unique and many other can be proposed to have a similar or
related spectrum. There are number of atomic schemes in literatures having
some similarities with our work. However a careful analysis is needed to
interpret these results in association with the Autler-Townes multiplet
spectra.

We begin with the discussion of correlation of the Autler-Townes multiplet
spectra with the existing literature concerning with the emission
spectroscopy. In reference \cite{Bibhas}, the control of spontaneous
emission is studied in a coherently driven four levels atomic system with an
excited decaying bare energy state. This scheme is an approximate version of
the Autler-Townes triplet spectrum \cite{Fazal}. Also, the work \cite%
{Bibhas01} presented by Bibhas and Prasanta in a N-type atom driven by two
coherent fields is also an approximated version of Autler-Townes triplet
spectrum with introduction of an additional effect of collisional
broadening. Extreme line-width narrowing and broadening is observed in all
the three spectral components agreeing with the Autler-Townes triplet
spectrum.

Moreover, Ref. \cite{Shujing} discussed the manipulations of absorption
spectrum via double controlled destructive (constructive) interference via
multiple routes to excitation in the four levels coherently driven atomic
system. Of course, they interpreted their results in connection with the EIT
and its utilization in logic gates, sensitive optical switches and quantum
coherence information storage due to the dispersive properties of the
medium. In their results two dark lines are always there which needs careful
interpretation regarding Autler-Townes multiplet spectroscopy. A $\Lambda $%
-type four levels atomic scheme was conducted experimentally by Chun-Liang
et al. \cite{Chun} for the absorption and dispersive properties of the
medium in the context of specific emission processes. They demonstrated that
double transparency windows with the controllable narrow central peak can be
observed due to the spontaneously generated coherence (SGC) without the need
of the rigorous condition of nearly degenerated levels. The
double-transparency windows are normally resulting from the double dark
lines in the absorption profile. These dark lines are traditionally traced
back to the interference mechanism of Fano type but is also the primary part
of Autler-Townes multiplet spectroscopy \cite{Fazal111} under ideal
conditions. Therefore, these results can be best interpreted using Fano type
interference mechanism as an integral part of Autler-Townes multiplet
spectroscopy. However, for large decay rates, the Weisskopf Wigner theory
may not be satisfied regarding the widths and also the Fano profile of the
dark lines of the spectra may not be significant. Next, the absorption and
consequently the dispersive properties of the medium is also discussed in
tripod type atom \cite{Hou}. Again here the destructive interference
mechanism in the excitation probabilities via three distinct paths to the
three excited dressed state also requires interpretation of the results in
the context of Autler-Townes triplet absorption spectroscopy. The two dark
lines appearing in the spectrum lead to two transparency windows at two
location on the frequency axis. The correlation of the Autler-Townes
multiplet schemes with the other related schemes in the context of
absorption spectra is broad in its range. However, all these studies are not
necessarily having spectral narrowing in all components. Further, to the
best of our knowledge only Autler-Townes triplet type absorption schemes may
be found in literature.

\section{Conclusion}

Generally, when the excited state is also coupled with a third level by
strong laser field through an allowed electric dipole transition then the
excitation signal is modified and it appears as two distinct components, a
pattern often termed the Autler-Townes doublet, honoring those who first
observed it \cite{Autler}. In Refs. \cite{Fazal,Fazal00} we extended their
idea to three distinct components (bright lines) in a spontaneous emission
spectrum by modifying their way of atom-field interaction and named in their
honor as Autler-Townes triplet spectrum.

We further proposed schemes for the generalization of Autler-Townes
spectroscopy from doublet and triplet to quartuplet, quintuplet and, suggest
linkages in sodium atom in which to display these spectra. We also
underlined the way how to extend to more higher ordered multiplet
spectroscopy. To understand the fundamentals of the processes ideally we
examined the Laplace transform for the statevector of these multi-state
systems subjected to steady coherent illumination in the rotating wave
approximation and Weisskopf-Wigner treatment of spontaneous emission as a
simple probability loss. We learned from the study of different
Autler-Townes multiplet schemes in which the single Lorentzian line shaped
spectrum can be divided into any number of spectral components depending on
the nature of the atom, nature of required fields and their interactions. In
this way we can split the decaying bare energy state into any number of
dressed energy state playing a vital role in generating the processes of
quantum interferences during the evolution of the systems. Further, the
widths of all the spectral components can be controlled through the relative
strength of the laser fields in a way obeying Weisskopf-Wigner theory. This
behavior is different from Autler-Townes doublet where no such an effect
exists through the strength of the driving fields. Furthermore, it is
experimentally very easy to control the strength of the laser fields as
these are externally controllable parameters. In this way an extreme
line-width narrowing can be achieved in all the spectral lines resulting
from the emission probabilities of all the dressed energy states of the bare
energy state for the atom in all the schemes under investigation.

The phenomenon of Autler-Townes multiplet spectroscopy can be realized
experimentally in alkali metal, for example, Na-atom \cite{Mirza}, and
molecule \cite{Xia,Ruth}. Further, if the interactions of the fields with
the atoms or molecules are consider in a cavity then it is certain to
include the velocity distribution and the decay terms due to the interaction
with the buffer gas. This may modify some phenomenon presented in this
paper; however, the underlying physics of the quantum interference is still
valid. Furthermore, if the experiment is performed in magneto-optical-trap
(MOT) \cite{MOT} where the atomic temperature can be decreased up to few
tents $\mu K$ then the Doppler broadening effect can be eliminated
significantly. Reliable quantitative analysis of the profiles of
Autler-Townes multiplet spectroscopy including the peaks height, widths and
spectral lines separation and dark line feature requires detailed modelling
of relevant atomic or molecular states with all possible decaying rates,
with inclusion of magnetic sub-degeneracy, Doppler shift distribution,
collisional broadening and many other decoherence processes.

The complexities occurred due to increase in the number of laser fields
interacting with the complex atomic systems are naturally inherited for the
conceptual foundation of the mechanisms of the interference involved in the
multiplet spectroscopy. However, in nature we are not limited to the domain
of simple atomic systems. In fact there are complex structured atoms,
molecules and clusters where the higher-ordered multiplet spectroscopy can
be realized theoretically and experimentally.

Our proposed schemes for this higher-ordered multiplet spectroscopy will
further open up new ways to explore, particularly, the physics of
spontaneous emission, stimulated absorption, their population dynamics,
multiwaves mixing and EIT while generally, it may be of interest to some
physicists and chemists working in the areas of laser spectroscopy, quantum
optics and nonlinear optics.

\appendix{}

\section{\textbf{Equations of Motion for Autler-Townes quartuplet
Spectroscopy}}

We derive the equations of motion for the probability amplitudes of the
state-vector using Schr{\small \"{O}}dinger equation and Weisskopf-Wigner
theory, which are given as%
\begin{align}
\overset{\cdot }{E}(t)& =-i\Omega _{o_{1}}G_{1}(t)-i\Omega _{o_{2}}G_{2}(t)-%
\frac{\Gamma }{2}E(t),  \label{A2} \\
\overset{\cdot }{G}_{1}(t)& =-i\Omega _{m_{1}}G_{3}(t)-i\Omega
_{o_{1}}^{\ast }E(t),  \label{A3} \\
\overset{\cdot }{G_{2}}(t)& =-i\Omega _{o_{2}}^{\ast }E(t)-i\Omega
_{m_{2}}G_{3}(t), \\
\overset{\cdot }{G}_{3}(t)& =-i\Omega _{m_{1}}^{\ast }G_{1}(t)-i\Omega
_{m_{2}}^{\ast }G_{2}(t),  \label{A4} \\
\overset{\cdot }{G}_{k}(t)& =-ig_{k}^{\ast }e^{-i\Delta _{k}t}E(t),
\label{A5}
\end{align}%
where $\Gamma $ is the decay rate from level $\left\vert a\right\rangle $ to
the ground level $\left\vert g\right\rangle $. Further, we assume that
initially, at time $t=0,$ the atom is prepared in the level $\left\vert
E\right\rangle $, therefore, $E(0)=1$ and $G_{1}(0)=G_{2}(0)=G_{3}(0)=G(0)=0$%
. In our system we only consider $\Omega _{o_{1}}$ as a complex quantity
i.e., $\left\vert \Omega _{o_{1}}\right\vert e^{i\varphi }$ whereas $\Omega
_{o_{2}}$, $\Omega _{m_{1}}$ and $\Omega _{m_{2}}$ are real i.e., $\Omega
_{o_{2}}=\left\vert \Omega _{o_{2}}\right\vert $, $\Omega
_{m_{1}}=\left\vert \Omega _{m_{2}}\right\vert $ and $\Omega
_{m_{3}}=\left\vert \Omega _{m_{2}}\right\vert $. We use Laplace transforms
to obtain expressions for the time-dependent amplitudes. Using Eq. (A5) we
then obtain the required asymptotic amplitude in the form

\begin{equation}
G_{k}(t\rightarrow \infty )=\frac{N\left( \Delta _{k}\right) }{D\left(
\Delta _{k}\right) },  \label{A6}
\end{equation}%
where%
\begin{equation}
N\left( \Delta _{k}\right) =g_{k}^{\ast }\Delta _{k}(\left\vert \Omega
_{m_{2}}\right\vert ^{2}+\left\vert \Omega _{m_{3}}\right\vert ^{2}-\Delta
_{k}^{2}),  \label{A7}
\end{equation}%
and
\begin{align}
D\left( \Delta _{k}\right) & =\Delta _{k}^{4}-\left( \left\vert \Omega
_{o_{2}}\right\vert ^{2}+\left\vert \Omega _{m_{1}}\right\vert
^{2}+\left\vert \Omega _{m_{2}}\right\vert ^{2}\right) \Delta
_{k}^{2}+\left\vert \Omega _{o_{2}}\right\vert ^{2}\left\vert \Omega
_{m_{2}}\right\vert ^{2}+\left\vert \Omega _{o_{1}}\right\vert \left\vert
\Omega _{o_{2}}\right\vert \left\vert \Omega _{m_{1}}\right\vert \left\vert
\Omega _{m_{2}}\right\vert \sin \varphi  \notag \\
& +i\frac{\Gamma }{2}\Delta _{k}(\left\vert \Omega _{m_{2}}\right\vert
^{2}+\left\vert \Omega _{m_{3}}\right\vert ^{2}-\Delta _{k}^{2}).  \label{A8}
\end{align}%
The asymptotic form of the state-vector is given by
\begin{subequations}
\begin{equation}
\left\vert \Psi (t\rightarrow \infty )\right\rangle
=\sum\limits_{k}G_{k}(t\longrightarrow \infty )\left\vert g\right\rangle
\left\vert 1_{k}\right\rangle .  \label{A9}
\end{equation}%
Using the above steady state-vector we find the spontaneous emission
spectrum proportional to

\end{subequations}
\begin{equation}
S(\Delta )\varpropto \left\vert G(\infty )\right\vert ^{2}  \label{A10}
\end{equation}%
where $G_{k}(t\rightarrow \infty )$ is replaced by $G(\infty )$ under the
consideration of the replacement of the discrete frequencies and the
detuning of the emission spectrum by their a continuum values using
Weisskopf-Wigner theory. Therefore apart from the proportionality factor,
the spontaneous emission spectrum is then given by Eq. (\ref{5})

Next, the factor $\digamma $ appearing in the Eq. (\ref{11}) is given by
\begin{align}
\digamma & =\xi _{1}^{3}(\xi _{3}-\xi _{4})(\xi _{2}^{2}+\xi _{3}\xi
_{4}-2\xi _{2}\xi _{4})+\xi _{2}^{3}[(\xi _{4}-\xi _{3})(\xi _{1}^{2}+\xi
_{3}\xi _{4})  \notag \\
& -\xi _{1}(\xi _{4}^{2}-\xi _{3}^{2})]+\xi _{3}^{3}(\xi _{4}-\xi _{1})[\xi
_{1}\xi _{4}-\xi _{2}(\xi _{4}-\xi _{2})]+\xi _{4}^{3}[(\xi _{3}-\xi _{1})
\notag \\
& (\xi _{1}^{2}-\xi _{1}\xi _{2}-\xi _{2}^{2})+\xi _{2}(\xi _{4}^{2}-\xi
_{1}^{2})]+\xi _{1}\xi _{2}\xi _{3}\xi _{4}[\xi _{1}(1+\xi _{4})-2\xi
_{3}\xi _{4}],  \label{A11}
\end{align}%
while the terms $\zeta _{i}$ $(i=1-4)$ appearing in the same equation are
\begin{align}
\zeta _{1}& =(\xi _{2}^{2}+\xi _{3}\xi _{4})(\xi _{4}-\xi _{3})-\xi _{2}(\xi
_{4}^{2}-\xi _{3}^{2}),  \label{A12} \\
\zeta _{2}& =(\xi _{3}^{2}+\xi _{4}\xi _{1})(\xi _{1}-\xi _{4})-\xi _{3}(\xi
_{1}^{2}-\xi _{4}^{2}),  \label{A13} \\
\zeta _{3}& =(\xi _{4}^{2}+\xi _{1}\xi _{2})(\xi _{2}-\xi _{1})-\xi _{4}(\xi
_{2}^{2}-\xi _{1}^{2}),  \label{A14} \\
\zeta _{4}& =(\xi _{1}^{2}+\xi _{2}\xi _{3})(\xi _{3}-\xi _{2})-\xi _{1}(\xi
_{3}^{2}-\xi _{2}^{2}),  \label{A15}
\end{align}%
respectively.

\section{\textbf{Equations of Motion for Autler-Townes Quintuplet
Spectroscopy}}

We derive the equations of motion for the probability amplitudes of the
state-vector (\ref{15}) using Schr$\mathtt{\ddot{o}}$dinger equation and
Weisskopf-Wigner theory, which are given as%
\begin{align}
\overset{\cdot }{E}(t)& =-i\Omega _{o_{1}}G_{1}(t)-i\Omega _{o_{2}}G_{2}(t)-%
\frac{\Gamma }{2}E(t),  \label{B1} \\
\overset{\cdot }{G}_{1}(t)& =-i\Omega _{m_{1}}G_{4}(t)-i\Omega
_{o_{1}}^{\ast }E(t),  \label{B2} \\
\overset{\cdot }{G}_{2}(t)& =-i\Omega _{m_{3}}G_{4}(t)-i\Omega
_{o_{2}}^{\ast }E(t)-i\Omega _{m_{1}}G_{3}(t),  \label{Q} \\
\overset{\cdot }{G}_{3}(t)& =-i\Omega _{m_{1}}^{\ast }G_{1}(t)-i\Omega
_{m_{2}}^{\ast }G_{2}(t),  \label{B4} \\
\overset{\cdot }{G}_{4}(t)& =-i\Omega _{m_{3}}^{\ast }G_{2}(t),  \label{S} \\
\overset{\cdot }{G}_{k}(t)& =-ig_{k}^{\ast }e^{-i\Delta _{k}t}E(t),
\label{B6}
\end{align}%
where $\Gamma $ is the decay rate from upper level $\left\vert
e\right\rangle $ to the ground level $\left\vert g\right\rangle $. Further,
we assume that initially, at time $t=0,$ the atom is prepared in the level $%
\left\vert e\right\rangle $, therefore, $E(0)=1$ and $%
G_{1}(0)=G_{2}(0)=G_{3}(0)=G_{4}(0)=G_{5}(0)=0$. Again we assume that $%
\Omega _{o_{1}}$ to be complex i.e., $\left\vert \Omega _{o_{1}}\right\vert
e^{i\varphi }$ however $\Omega _{o_{2}}$, $\Omega _{m_{1}}$, $\Omega
_{m_{2}} $ and $\Omega _{m_{3}}$ to be real i.e., $\Omega
_{o_{2}}=\left\vert \Omega _{o_{2}}\right\vert $, $\Omega
_{m_{1}}=\left\vert \Omega _{m_{1}}\right\vert $,$\Omega _{m_{2}}$ $%
=\left\vert \Omega _{m_{2}}\right\vert $ and $\Omega _{m_{3}}=\left\vert
\Omega _{m_{3}}\right\vert $. We use Laplace transform method to evaluate
the steady state expression for the probability amplitudes $%
G_{k}(t\rightarrow \infty )$, as%
\begin{equation}
G_{k}(t\rightarrow \infty )=\frac{N\left( \Delta _{k}\right) }{D\left(
\Delta _{k}\right) },  \label{B7}
\end{equation}%
where%
\begin{equation}
N\left( \Delta _{k}\right) =g_{k}^{\ast }\left[ \Delta _{k}^{4}-\Delta
_{k}^{2}\left( \left\vert \Omega _{m_{1}}\right\vert ^{2}+\left\vert \Omega
_{m_{2}}\right\vert ^{2}+\left\vert \Omega _{m_{3}}\right\vert ^{2}\right)
+\left\vert \Omega _{m_{2}}\right\vert ^{2}\left\vert \Omega
_{m_{3}}\right\vert ^{2}\right] ,  \label{B8}
\end{equation}%
and%
\begin{align}
D\left( \Delta _{k}\right) & =-\Delta _{k}^{5}+\Delta _{k}^{3}\left(
\left\vert \Omega _{o_{1}}\right\vert ^{2}+\left\vert \Omega
_{m_{1}}\right\vert ^{2}+\left\vert \Omega _{m_{2}}\right\vert
^{2}+\left\vert \Omega _{m_{3}}\right\vert ^{2}+\left\vert \Omega
_{o_{2}}\right\vert ^{2}\right)  \notag \\
& -\Delta _{k}(\left\vert \Omega _{o_{1}}\right\vert ^{2}\left\vert \Omega
_{3}\right\vert ^{2}+\left\vert \Omega _{m_{1}}\right\vert ^{2}\left\vert
\Omega _{o_{2}}\right\vert ^{2}+\left\vert \Omega _{m_{2}}\right\vert
^{2}\left\vert \Omega _{m_{3}}\right\vert ^{2}+\left\vert \Omega
_{m_{3}}\right\vert ^{2}\left\vert \Omega _{o_{2}}\right\vert ^{2}  \notag \\
& -2\left\vert \Omega _{o_{1}}\right\vert \left\vert \Omega
_{m_{1}}\right\vert \left\vert \Omega _{m_{2}}\right\vert \left\vert \Omega
_{m_{2}}\right\vert \left\vert \Omega _{o_{2}}\right\vert \cos \varphi )
\notag \\
& +i\frac{\Gamma }{2}(\Delta _{k}^{4}-\Delta _{k}^{2}\left( \left\vert
\Omega _{m_{1}}\right\vert ^{2}+\left\vert \Omega _{m_{2}}\right\vert
^{2}+\left\vert \Omega _{m_{3}}\right\vert ^{2}\right) +\left\vert \Omega
_{m_{2}}\right\vert ^{2}\left\vert \Omega _{m_{3}}\right\vert ^{2}).
\label{B9}
\end{align}%
Further, in Eqs. (\ref{30}), the factor $\Lambda $ and the factors $\kappa
_{i}$ $\left( i=1-5\right) $ are given by

\begin{align}
\Lambda & =(\xi _{1}-\xi _{2})(\xi _{1}-\xi _{3})(\xi _{1}-\xi _{4})(\xi
_{1}-\xi _{5})(\xi _{2}-\xi _{3})  \notag \\
& (\xi _{2}-\xi _{4})(\xi _{2}-\xi _{5})(\xi _{3}-\xi _{4})(\xi _{3}-\xi
_{5})(\xi _{4}-\xi _{5}),  \label{B10}
\end{align}%
and%
\begin{align}
\kappa _{1}& =(\xi _{2}-\xi _{3})(\xi _{2}-\xi _{4})(\xi _{2}-\xi _{5})(\xi
_{3}-\xi _{4})(\xi _{3}-\xi _{5})(\xi _{4}-\xi _{5}),  \label{B11} \\
\kappa _{2}& =(\xi _{1}-\xi _{3})(\xi _{1}-\xi _{4})(\xi _{1}-\xi _{5})(\xi
_{3}-\xi _{4})(\xi _{3}-\xi _{5})(\xi _{4}-\xi _{5}),  \label{B12} \\
\kappa _{3}& =(\xi _{1}-\xi _{2})(\xi _{1}-\xi _{4})(\xi _{1}-\xi _{5})(\xi
_{2}-\xi _{4})(\xi _{2}-\xi _{5})(\xi _{4}-\xi _{5}),  \label{B13} \\
\kappa _{4}& =(\xi _{1}-\xi _{2})(\xi _{1}-\xi _{3})(\xi _{1}-\xi _{5})(\xi
_{2}-\xi _{3})(\xi _{2}-\xi _{5})(\xi _{3}-\xi _{5}),  \label{B14} \\
\kappa _{5}& =(\xi _{1}-\xi _{2})(\xi _{1}-\xi _{3})(\xi _{1}-\xi _{4})(\xi
_{2}-\xi _{3})(\xi _{2}-\xi _{4})(\xi _{3}-\xi _{4}).  \label{B15}
\end{align}%
respectively.


\begin{thebibliography}{99}
\bibitem{Book1} See, for example, M. O. Scully and M. S. Zubairy, \textit{%
Quantum Optics}\textrm{\ }(Cambridge University Press, Cambridge, 1997); W.
H. Louisell, \textit{Quantum Statistical Properties of Radiation }(Wiley,
New York, 1973).

\bibitem{Autler} S. H. Autler and C. H. Townes, Phys. Rev. \textbf{100,}
4963 (1955).

\bibitem{Fano} U. Fano, "Effect of configuration interaction on intensities and phase shifts'
Phys. Rev. \textbf{124}, 1866 (1961); U. Fano and J.
W. Cooper., Phys. Rev. A \textbf{340}, 441 (1968); P. L. Knight and P. W.
Milonni, Phys. Rep. \textbf{66,} 23 (1980); Plastina Francesco and Piperno
Francesco, Phys. Lett. A \textbf{236}, 16 (1997).

\bibitem{Agarwal} G. S. Agarwal, in \textit{Quantum Optics Statistical
Theories of Spontaneous Emission and Their Relation to Other Approaches},
edited by G. H\"{o}hler et al., Springer Tracts in Modern Physics Vol.
\textbf{70} (Springer-Verlag, Berlin, 1974).

\bibitem{Agassi} D. Agassi, "Spontaneous radiative decay of a continuum" Phys. Rev. A \textbf{30}, 2449 (1984).

\bibitem{Zhu-Narducci} S.-Y. Zhu, L. M. Narducci and M. O. Scully,
"Qunatum mechanical interference effects in the spontaneous-emission spectrum of a driven atom" Phys.
Rev. A \textbf{52,} 4791 (1995).

\bibitem{Knight01} E. Paspalakis, C. H. Keitel and P. L. Knight,
"Flourescenec control through multiple interference mechanism" Phys. Rev.
A \textbf{58,} 4868 (1998).

\bibitem{Physik} T. H\"{a}nsch, R. Keil, A. Schabert, Ch. Schmelzer and P.
Toschek\textit{,} "Interaction of laser light waves by dynamic
Stark splitting", Z. Physik \textbf{226}, 293 (1969).

\bibitem{Keil} A. Schabert, R. Keil and P. E. Toscheck, Opt. Commun. \textbf{%
13}, 265 (1975).

\bibitem{Keil00} A. Schabert, R. Keil and P. E. Toscheck,
"Dynamic Stark effect of an optical line observed by cross saturated absorbtion", Appl. Phys.
\textbf{6}, 181 (1975).

\bibitem{Pic} J. L. Picque and J. Pinard, J. Phys. B\textbf{\ 9}, 177 (1976).

\bibitem{Ph} P. Cahuzac and R. Vetter,
"Observation of Autler-Townes effect on infrared laser transitions of Xenon", Phys. Rev. A \textbf{14}, 270 (1976).

\bibitem{Bjor} J. E. Bjorkholm and P. F. Liao, Opt. Commun. \textbf{21}, 132
(77).

\bibitem{Gray} H. R. Gray and C. R. Stroud, Jr Opt. Commun. \textbf{25}, 359
(1978).

\bibitem{Delsart} C. Delsart, J. C. Keller and V. P. Kaftandjian, J. Phys.
(Paris) \textbf{42}, 529 (1981).

\bibitem{Fisk} P. T. H. Fisk, H.-A. Bachor and R. J. Sandeman,"Investigation
of the dynamic Stark effect in a $J\rightarrow 0 \rightarrow1$
three-level system. I. Experiment" Phys. Rev. A \textbf{33}, 2418
(1986).

\bibitem{Autler-Zubairy} M. S. Zubairy, "Qunatum state measurement via Autler-Townes spectroscopy" Phys. Lett. A \textbf{222, }91
(1996).

\bibitem{Wigner-Zubairy} M. Mehmoudi, H. Tajalli and M. S. Zubairy,
"Measurement of Wigner function of a cavity field via Autler-Townes spectroscopy", Jr. Opt.
B: Quant. Semiclass. Opt. \textbf{2}, 315 (2000).

\bibitem{Zubairy-Manzoor} M. Ikram and M. S. Zubairy,
"Reconstruction of an entangled state in a cavity via
Autler-Townes spectroscopy", Phys. Rev\textit{. }A \textbf{65},
044305 (2002).

\bibitem{Herkommer} A. M. Herkommer, W. P.\ Schleich, and M. S.
Zubairy, "Autler-Townes microscopy on a single atom", Jr. Mod.
Opt. \textbf{44}, 2507 (1997).

\bibitem{Fazal} F. Ghafoor, S. Qamar, S.- Y. Zhu, and M. S. Zubairy, "Autler-Townes triplet spectroscopy" Opt.
Commun. \textbf{273}, 464 (2007).

\bibitem{Fazal00} F. Ghafoor, "Autler-Townes triplet spectroscopy: an nalytical approach", Opt. Commun. \textbf{284}, 1913 (2011).

\bibitem{Ghafoor} F. Ghafoor, S. Qamar and M. S. Zubairy, "Atom localization via phase and amplitude control of the driving fields", Phys. Rev. A
\textbf{65}, 043819 (2002).

\bibitem{Ghafoor00} F. Ghafoor, "Subwavelength atom localization via quantum coherence"' Phys. Rev. A \textbf{84}, 063849 (2011).

\bibitem{F-Ghafoor} F. Ghafoor, S.-Y. Zhu, and M. S. Zubairy, "Amplitude and phase control of spontaneous emission", Phys. Rev. A
\textbf{62}, 13811 (2000).

\bibitem{Sehrai} M. Sahrai, H. Tajalli, K. T. Kapale and M. S.
Zubairy, "Subwavelength atom localization via amplitude and phase
control of the absorption spectrum ", Phys. Rev. A \textbf{72},
013820 (2005).

\bibitem{Sehrai1} M. Sahrai, H. Tajalli, K. T. Kapale and M. S.
Zubairy, "Tunable phase control for subluminal to superluminal
light propagation",  Phys. Rev. A \textbf{70}, 023813 (2004).

\bibitem{Mirza} A. B. Mirza and S. Singh, "Wave-vector mismatch effects in electromagnetically induced transparency in $Y$-type systems", Phys. Rev. A \textbf{85}, 053837
(2012).

\bibitem{Xia} H.-R. Xia, C.-Y. Ye, and S.-Y. Zhu, "Experimental Observation of Spontaneous Emission Cancellation", Phys. Rev. Lett. \textbf{77%
}, 1032 (1996)\nolinebreak .

\bibitem{Zanch} Z. Zuo, J. Sung, X. Liu, Q. Jiang, G. Fu, L-A. Wu and P. Fu,
Phys. Rev. Lett. \textbf{97}, 193904 (2004).

\bibitem{Yanp} Y. Zhang, B. Anderson and M. Xiao, J. Phys. B, \textbf{41},
045502 (2008).

\bibitem{Zhiq} Z. Nie, H. Z. P. Li, Y. Yang, Y. Zhang and M. Xiao, Phys.
Rev. A, Iss. \textbf{77}, 6 (2008).

\bibitem{Yig} Y. Du, Y. Zhang, C. Zuo, C. Li, Z. Nie, H. Zheng, M. Shi, R.
Wang, J. Song, K. Lu and M. Xiao, "Generalized n-Photon Resonant
$2n$-Wave Mixing in an $(n + 1)$-Level System with Phase-Conjugate
Geometry", Phys. Rev. A, \textbf{79}, 063839 (2009).

\bibitem{Chang} C. Li, Y. Zhang, Z. Nie, H. Zheng, C. Zuo, Y. Du, J. Song,
K. Lu, and C. Gan, Opt. Commun. \textbf{283}, 29182928 (2010).

\bibitem{Yanp01} Y. Zhang, P. Li, H. Zheng, Z. Wang, H, Chen, C. Li, R.
Zhang and M. Xiao, Opt. Exp, \textbf{19}, 7769 (2011).

\bibitem{Chang01} C. Li, Y. Zhang, Z. Wang, Y. Li, Z. Nie, Y. Zhao, R. Wang,
C. Yuan and K. Lu, Opt. Commun. \textbf{284}, 1379 (2011).

\bibitem{Paspalakis} E. Paspalakis and P. L. Knight, "Phase Control of Spontaneous Emission", Phys. Rev. Lett.
\textbf{81}, 293 (1998).

\bibitem{Zhu-Scully} S.-Y. Zhu and M. O. Scully, Phys. Rev. Lett. \textbf{76}%
, 388 (1996).

\bibitem{Fry} E. S. Fry, X. Li, D. Nikonov, G. G. Padmabandu, M. O. Scully,
A. V. Smith, F. K. Tittel, C. Wang, S. R. Wilkinson and S-Y. Zhu,
"Atomic Coherence Effects within the Sodium D line: Lasing without
Inversion via Population Trapping", Phys. Rev. Lett. \textbf{70},
3235 (1993);G. G. Padmabandu, X. Li, C. Su, E. S. Fry, D. Nikonov,
S-Y. Zhu, G. M. Meyer and M. O. Scully, Quantum Semiclass. Opt.
\textbf{6}, 261, (1994).

\bibitem{Fry01} G. M. Meyer, U. W. Rathe, M. Graf, S-Y. Zhu, E. S. Fry, M.
O. Scully, G. H. Herling and L. M. Narducci, Quantum Semiclass. Opt. \textbf{%
6}, 231, (1994;D. E. Nikonov, U. W. Rathe, M. O. Scully, Shi-Yao. Zhu, E. S.
Fry, X. Li, G. G. Padmabandu, M. Plschhauer, Quantum Semiclass. Opt. \textbf{%
6}, 245, (1994).

\bibitem{Barber} Z. W Barber, C. W. Hoyt, C. W. Oates, L. Hollberg, A. V.
Taichenachev and V. I Yudin, arXive: physics/0512084v2 [physics.atom-ph] 2
march 2006.

\bibitem{copy} C. Ding, J. Li, R. Yu, X. Hao, and Y. Wu, Optics Express,
\textbf{20}, 7871 (2012).

\bibitem{Mart} M. A. G. Martinez, P. R. Herezfeld, C. Sannuels, L. M.
Narducci and C. H. Keitel, "Quantum interference effects in
spontaneous atomic emission: Dependence of the resonance
?uorescence spectrum on the phase of the driving field", Phys.
Rev. A \textbf{55,} 4483 (1997).\nolinebreak \allowbreak

\bibitem{Bibhas} B. K. Butta and P. K. Mahapatra, Opt. Commun. \textbf{284},
594 (2009).

\bibitem{Bibhas01} B. K. Butta and P. K. Mahapatra, "Control of the spontaneous emission in a Driven
$N$-type by dynamically induced quantum interference", Phys, Scr. \textbf{79},
065402 (2009).

\bibitem{Shujing} S. Li, X. Yang, X. Cao, \ C. Xie and H. Wang, "Two electromagnetically induced transparency
windows and an enhanced electromagnetically induced transparency
signal in a four-level tripod atomic system", J. Phys. B,
\textbf{40}, 3211 (2007).

\bibitem{Chun} C. L. Wang, A. J. Li, X. Y. Zhou, Z. H. Kang, J. Yun and J. Y
Gao, "Passively mode-locked $Yb:YAG$ thin-disk laser with pulse
energies exceeding 13 $\mu$J by use of an 2 Joerg Neuhaus, 1 1,2,
active multipass geometry", Optics Letters, \textbf{33}, 7 (2008).

\bibitem{Fazal111} Absorption spectra are more complex than the emission
spectra having varous other type of results associated with the dispersion
profile of a system under certain general conditions. However under certain
limiting conditions we may get the results for the required phase sensitive
spectroscopy (to be submitted).

\bibitem{Hou} B. P. Hou, S. J. Wang, W. L. Yu and W. L. Sun, "Effects of vacuum-induced coherence on dispersion
and absorption properties in a tripod-scheme atomic system", J.
Phys. B, \textbf{39}, 2335 (2006).

\bibitem{Ruth} R. Garcia-Fernandez, A. Ekers, J. Klavins, L. P. Yatsenko, N.
N. Bezuglov, B. W. Shore and K. Bergmann, "Autler-Townes effect in
a sodium molecular-ladder scheme", Phys, Rev. A \textbf{71},
023401 (2005).\allowbreak

\bibitem{MOT} E. L. Raab, M. Prentiss, A. Cable, S. Chu and D. E. Pritchard,
"Trapping of Neutral Sodium Atoms with Radiation Pressure", Phys.
Rev. Lett. \textbf{59}, 2631 (1987).\allowbreak \nolinebreak
\newpage
\end{thebibliography}
\end{document}